\documentclass[aps,prd,superscriptaddress,amsmath,groupedaddress,twocolumn,showpacs]{revtex4}
\usepackage{amssymb,graphicx}

\newcounter{gplfmfgraph}

\newcommand{\smalltext}{\footnotesize}
\newcommand{\be}{\begin{equation}}
\newcommand{\ee}{\end{equation}}
\newcommand{\bea}{\begin{eqnarray}}
\newcommand{\eea}{\end{eqnarray}}
\newcommand{\bc}{\begin{center}}
\newcommand{\ec}{\end{center}}
\newcommand{\bi}{\begin{itemize}}
\newcommand{\ei}{\end{itemize}}
\newcommand{\bis}{\begin{itemize} \smalltext}
\newcommand{\eis}{\end{itemize}}
\newcommand{\biss}{\begin{itemize} \smalltext}
\newcommand{\eiss}{\end{itemize}}
\newcommand{\bfr}{\begin{frame}}
\newcommand{\efr}{\end{frame}}
\newcommand{\order}{{\cal O}}

\newcommand{\e}{{\rm e}}

\newcommand{\psib}{\overline{\psi}}
\newcommand{\ihalf}{\mbox{$\frac{i}{2}$}}

\newcommand{\Tr}{{\rm\,Tr\,}}

\newcommand{\Dv}{\mathbf{D}}
\newcommand{\xv}{{\bf x}}

\newcommand{\pv}{{\bf p}}

\newcommand{\Ev}{\mathbf{E}}
\newcommand{\Bv}{\mathbf{B}}

\newcommand{\B}{{\cal B}}
\newcommand{\Bz}{{\cal B}_\zeta}

\newlength{\boxitlen}

\newlength{\shrinkboxlen}

\def\lag{{\cal L}}    

\def\quarter{\mbox{$\frac{1}{4}$}}

\newcommand{\eq}[1]{Eq.~(\ref{#1})}
\newcommand{\nl}{\nonumber \\}
\newcommand{\bearray}{\begin{eqnarray}}
\newcommand{\eearray}{\end{eqnarray}}

\newcommand{\omittext}[1]{}

\newlength{\minuslength}
\newlength{\digitlength}
\newcommand{\action}{{\cal S}}
\newcommand{\Parity}{\mathrm{P}}

\newcommand{\F}{{\cal F}}
\newcommand{\Ft}{{\cal F}^\mathrm{ASQTAD}}
\newcommand{\Ftt}{{\cal F}^\mathrm{HISQ}}
\newcommand{\chib}{\overline{\chi}}

\newcommand{\tot}{\mathrm{tot}}
\newcommand{\dprod}{\otimes}
\newcommand{\cJ}[2]{{\cal J}^{(#1)}_{#2}}

\newcommand{\deltax}{{\delta x}}

\begin{document}

\title{Highly Improved Staggered Quarks on the Lattice,\\ with 
Applications to Charm Physics.}

\author{E. Follana}
\email{e.follana@physics.gla.ac.uk}
\affiliation{Department of Physics and Astronomy, University of 
Glasgow, Glasgow, UK}
\author{Q. Mason}
\affiliation{Dept. Applied Maths and Theoretical Physics, Cambridge 
University, Cambridge, UK}
\author{C. Davies}
\affiliation{Department of Physics and Astronomy, University of 
Glasgow, Glasgow, UK}
\author{K. Hornbostel}
\affiliation{Southern Methodist University, Dallas, Texas 75275, USA}
\author{G. P. Lepage}
\affiliation{Laboratory for Elementary-Particle Physics, Cornell 
University, Ithaca, NY 14853, USA}
\author{J. Shigemitsu}
\affiliation{Physics Department, The Ohio State University, Columbus, 
Ohio 43210, USA}
\author{H. Trottier}
\affiliation{Physics Department, Simon Fraser University, Vancouver, 
British Columbia, Canada}
\author{K. Wong}
\affiliation{Department of Physics and Astronomy, University of 
Glasgow, Glasgow, UK}
\collaboration{HPQCD, UKQCD}
\noaffiliation

\date{\today}

\begin{abstract}
We use perturbative Symanzik improvement to create a new 
staggered-quark action (HISQ) that has greatly reduced one-loop 
taste-exchange errors, no tree-level order~$a^2$ errors, and no 
tree-level order~$(am)^4$ errors to leading order in the quark's 
velocity~$v/c$. We demonstrate with simulations that the resulting 
action has taste-exchange interactions that are at least 3--4 times 
smaller than the widely used ASQTAD action. We show how to estimate 
errors due to taste exchange by comparing ASQTAD and HISQ simulations, 
and demonstrate with simulations that such errors are no more than 1\% 
when HISQ is used for light quarks at lattice spacings of $1/10$\,fm or 
less. The suppression of $(am)^4$~errors also makes HISQ the most 
accurate discretization currently available for simulating $c$~quarks. 
We demonstrate this in a new analysis of the $\psi-\eta_c$~mass 
splitting using the HISQ action on lattices where $am_c=0.43$ and~0.66, 
with full-QCD gluon configurations (from MILC). We obtain a result 
of~111(5)\,MeV which compares well with experiment. We discuss 
applications of this formalism to $D$~physics and present our first 
high-precision results for $D_s$~mesons.
\end{abstract}


\maketitle

\section{Introduction}
The reintroduction of the staggered-quark discretization in recent 
years has transformed lattice quantum chromodynamics, making accurate 
calculations of a wide variety of important nonperturbative quantities 
possible for the first time in the history of the strong 
interaction~\cite{ratio-paper,fk-fpi,alphas-paper,fd,mb-paper}. 
Staggered quarks were introduced thirty years ago, but unusually large 
discretization errors, proportional to $a^2$ where $a$~is the lattice 
spacing, made them useless for accurate simulations. It was only in the 
late 1990s that we discovered how to remove the leading errors, and the 
result was one of the most accurate discretizations in use today. 
Staggered quarks are faster to simulate than other discretizations, and 
as a result have allowed us for the first time to incorporate (nearly) 
realistic light-quark vacuum polarization into our simulations. This 
makes high-precision simulations, with errors of order a few percent, 
possible for the first time (see~\cite{ratio-paper} for a more detailed 
discussion). In this paper we present a new discretization that is 
substantially more accurate than the improved discretization currently 
in use. With this new formalism accurate simulations will be possible 
at even larger lattice spacings, further reducing simulation costs.

The $\order(a^2)$~discretization errors in staggered quarks have two 
sources. One is the usual error associated with discretizing the 
derivatives in the quark action. The correction for this error is 
standard and was known in the 1980s\,\cite{naik-paper}. The second 
source was missed for almost a decade. It results from an unusual 
property of the staggered-quark discretization: the lattice quark field 
creates four identical flavors or tastes of quark rather than one. (We 
refer to these unphysical flavors as ``tastes'' to avoid confusion with 
the usual quark flavors, which are not identical since quarks of 
different flavor have different masses.) The missing $a^2$~error was 
associated with taste-exchange interactions, where taste is transferred 
from one quark line to another in quark-quark scattering. The generic 
correction for this type of error involves adding four-quark operators 
to the discretized action.

Taste is unphysical. Instead of one~$\pi^+$, for example, one has 
sixteen. Taste is easily removed in simulations provided that different 
tastes are exactly equivalent, which they are, and provided there are 
no interactions that mix hadrons of different taste. Taste-exchange 
interactions, however, cause such mixing, and therefore it is crucial 
that we understand such interactions and suppress them. Furthermore, 
past experience suggests that the errors due to residual taste exchange 
are the largest remaining $a^2$~errors in current simulations.

The corrections to the lattice action that suppress taste exchange were 
missed initially because four-quark operators are not usually needed to 
correct discretization errors in lowest order perturbation theory (that 
is, at tree level). They were first discovered empirically from 
simulations in which the gluon fields in the quark action were replaced 
by smeared fields, which happen to suppress taste-changing quark-quark 
scattering amplitudes\,\cite{doug-papers}.  This discovery led to a 
proper understanding of the taste-exchange 
mechanism\,\cite{lepage1-paper,sinclair-paper}, and to the first 
correct analysis of the $a^2$~errors in staggered-quark 
formalisms\,\cite{lepage1-paper,lepage2-paper}. The resulting 
``ASQTAD'' quark action has provided the basis for almost all 
simulations to date that include (nearly) realistic light-quark vacuum 
polarization.

While it is possible to remove all tree-level taste exchange by 
smearing the gluon fields appropriately, higher-order corrections can 
only be removed by adding four-quark operators. Recent numerical 
experiments suggest, however, that even these corrections can be 
significantly suppressed through additional smearing\,\cite{hyp-paper}. 
These studies, while very important, were limited in two ways. First, 
they used the mass splittings between the sixteen tastes of pion as a 
probe for taste exchange; there are many taste-exchange interactions 
that do not affect this spectrum. Second, smearing the gluon fields 
introduces other types of $\order(a^2)$~error which undo the advantage 
of an improved discretization (that is, accuracy at large values 
of~$a$).

In this paper we present the first rigorous analysis of one-loop taste 
exchange. Based upon this analysis we develop a simple smearing scheme 
that suppresses one-loop corrections by an order of magnitude on 
average. Our smearing scheme is simpler than that in~\cite{hyp-paper} 
and, unlike that scheme, it does not introduce new $a^2$~errors. The 
result is a new discretization for highly-improved staggered quarks 
which we refer to by the acronym~HISQ.

An added advantage of a highly improved action is that it can be used 
to simulate $c$~quarks. Heavy quarks are difficult to simulate using 
standard discretizations because discretization errors are large unless 
$am\ll1$, where $a$~is the lattice spacing and $m$~the quark mass. To 
simulate $b$~quarks, for example, one would require lattice spacings 
substantially smaller than $1/20$\,fm\,---\,much too small to be 
practical today. Consequently high-quality simulations of $b$~quarks 
rely upon rigorously defined effective field theories, like 
nonrelativistic QCD\,\cite{nrqcd}, that remove the rest mass from the 
quark's energy. While this approach works very well for $b$~quarks, it 
is less successful for $c$~quarks because the quark's mass is much 
smaller, and consequently $c$~quarks are much less nonrelativistic. The 
smaller quark mass, however, means that $am_c\approx 1/2$ for lattice 
spacings of order $1/10$\,fm, which are typical of simulations today. 
As we demonstrate in this paper, we can obtain few-percent accuracy for 
charm quarks when $am_c\approx1/2$ provided we use a highly improved 
relativistic discretization like~HISQ.

In Section~II we review the (perturbative) origins of taste-exchange 
interactions in staggered quarks, and their removal at lowest order 
(tree level) in perturbation theory, using the Symanzik improvement 
procedure. We also extend the traditional Symanzik analysis to cancel 
all $(am)^4$ errors to leading order in the quark's velocity, $v/c$ (in 
units of the speed of light~$c$). The additional correction needed is 
negligible for light quarks, but significantly enhances the precision 
of $c$-quark simulations.

In Section~III, we extend our analysis of taste exchange to one-loop 
order, and show how to suppress such contributions by approximately an 
order of magnitude, thereby effectively removing one-loop taste 
exchange from the theory. The resulting quark action (HISQ) is the 
first staggered-quark discretization  that is both free of lattice 
artifacts through~$\order(a^2)$ and effectively free of all taste 
exchange through one-loop order. We also identify the order~$\alpha_s 
(am)^2$ corrections to the action that are important for high-precision 
$c$~physics.

We illustrate the utility of our new formalism, in Section~IV, first by 
examining its effect on taste splittings in the pion spectrum. We also 
show how to directly measure the size of taste-exchange interactions in 
simulations, by comparing ASQTAD and HISQ simulations. There has been 
much discussion recently about the formal significance of 
taste-exchange interactions~\cite{creutz-paper-etc}. With our procedure 
it is possible to directly measure the effect of these interactions on 
physical quantities. We demonstrate the procedure by showing that 
residual taste-exchange interactions in HISQ contribute less than~1\% 
at current lattice spacings to light-quark quantities such as meson 
masses and meson decay constants.

We report on a new simulation of $c$~quarks and charmonium using the 
HISQ formalism in Section~V. This is the most severe test of the HISQ 
formalism since $am_c\approx1/2$ for the lattice spacings we use, but 
we show that the formalism delivers results that are accurate to a 
couple of percent, making it the most accurate formalism there is for 
simulating $c$~quarks. We demonstrate this in a new high-precision 
determination of the $\psi$--$\eta_c$~splitting. This analysis confirms 
our expectation that the largest $a^2$~errors in the ASQTAD action are 
associated with taste exchange, and we show that these are greatly 
reduced by using the HISQ action instead.

The moderately heavy smearing used in the HISQ formalism complicates 
unquenched simulations. In Section~VI, we discuss how these 
complications can be dealt with. Finally, in Section~VII, we summarize 
our results and discuss their relevance to future work. Here we also 
present our first $D$~physics results from the HISQ action.

Much of our discussion is framed in terms of the ``naive'' 
discretization for quarks, which is conceptually equivalent to the 
staggered-quark discretization (see 
Appendix~\ref{about-staggered-quarks}). We usually find naive quarks to 
be more intuitive in analytic work than staggered quarks, while the 
latter formalism is definitely the more useful one for numerical work. 
We outline the formal connection between the naive-quark and 
staggered-quark formalisms in a series of detailed Appendices. We use 
these results in our one-loop analysis of taste-exchange interactions, 
which is described in detail in Appendix~\ref{app-oneloop}.

We used several sets of gluon configurations in this paper. The 
parameters for these various sets are summarized in 
Table~\ref{config-table}.

\begin{table}
	\begin{center}
		\begin{tabular}{ccccccc}
			\hline\hline
			& $a$ (fm) & Gluon Action & $am_s$ & $am_{u/d}$ & $L/a$\\ \hline
			UKQCD & $1/10$ & Wilson  & -- & -- & 16 \\
			MILC & $1/8$ & L\"uscher-Weisz  & $0.05$ & 0.01 & 20 \\
			MILC & $1/11$& L\"uscher-Weisz  & $0.03$ & $0.006$ & 28 \\
			\hline\hline
		\end{tabular}
	\end{center}
	\caption{Gluon configurations used in this paper with information 
about
	the collaboration that produced them,
	the gluon action used (the unimproved quenched Wilson action
	or the Symanzik-improved L\"uscher-Weisz
	action for full $n_f=3$ QCD), the $s$
	and $u=d$~quark masses used in vacuum polarization,
	and the spatial size of the lattice. The first set of configurations 
is
	described in~\cite{ukqcd-w}, while the others are discussed in detail
	in~\cite{milc-configs,ratio-paper}.}
	\label{config-table}
\end{table}

\section{Symanzik Improvement for Naive/Staggered Quarks}
\subsection{Naive Quarks, Doubling, and Taste-Changing Interactions}
We begin our review of staggered quarks by examining the formally 
equivalent ``naive'' discretization of the quark action (see 
Appendix~\ref{about-staggered-quarks}):
\be\label{naive-quark-action}
\action = \sum_{x}\, \psib(x)\,(\gamma\cdot\Delta(U) +m_0)\,\psi(x).
\ee
where $\Delta_\mu$ is a discrete version of the covariant derivative,
\be
\begin{split}
\Delta_\mu(U)&\,\psi(x) \equiv\\
&\frac{1}{2a}\left(U_\mu(x)\,\psi(x+a\hat{\mu})
-U^{\dagger}_\mu(x-\hat{\mu})\,\psi(x-a\hat{\mu})\right),
\end{split}
\label{der}\ee
$U_\mu(x)$ is the gluon link-field, $a$ the lattice spacing, and $m_0$ 
is the bare quark mass. The gamma matrices are hermitian,
\be \label{gamma}
\gamma_\mu^\dagger = \gamma_\mu \quad\quad \gamma_\mu^2 = 1 \quad\quad
\{\gamma_\mu,\gamma_\nu\}=2\,\delta_{\mu\nu},
\ee
where indices $\mu$ and $\nu$ run over~0\ldots3.
A complete set of sixteen spinor matrices can be labeled by 
four-component vectors~$n_\mu$ consisting of~0s and~1s (i.e., 
$n_\mu\in\mathbb{Z}_2$):
\be
\gamma_n \,\equiv\, \prod_{\mu=0}^3(\gamma_\mu)^{n_\mu}.
\ee
A set of useful gamma-matrix properties is presented in 
Appendix~\ref{app-gamma}.

The naive-quark action has an exact ``doubling'' symmetry under the
transformation:
\bea
\psi(x)\quad \to \quad \tilde{\psi}(x) &\equiv& \gamma_5\gamma_\rho\,
(-1)^{x_\rho/a}\,\psi(x) \nl
&=& \gamma_{5}\gamma_\rho\,\exp(i\,x_\rho\pi/a)\,\psi(x).
\eea
Thus any low energy-momentum mode, $\psi(x)$, of the theory is 
equivalent to
another mode, $\tilde{\psi}(x)$, that has momentum $p_\rho \approx
\pi/a$, the maximum allowed on the lattice. This new mode is one of the
``doublers'' of  the naive
quark action. The doubling transformation can be
applied successively in two or more directions; the general
transformation is
\be \label{dsym}
\psi(x) \,\to\, \B_\zeta(x)\,\psi(x) \quad\quad
\psib(x) \,\to\, \psib(x)\,\Bz^\dagger(x)
\ee
where
\bea
\Bz(x) &\equiv& \gamma_{\,\overline{\zeta}}\,(-1)^{\zeta\cdot x/a} \nl
&\propto&
\prod_\rho\,\left(\gamma_5\gamma_\rho\right)^{\zeta_\rho}\,\,
\exp(i\,x\cdot\zeta\,\pi/a),\label{B}
\eea
and $\zeta$ is a vector with $\zeta_\mu\in\mathbb{Z}_2$, while 
$\overline\zeta$ is ``conjugate'' to~$\zeta$ (see 
Appendix~\ref{app-gamma}):
\begin{equation}
	\overline\zeta_\mu \equiv \sum_{\nu\ne\mu} \zeta_\nu \bmod 2
\end{equation}
Consequently there are 15~doublers in all (in four dimensions), which 
we label with the fifteen different nonzero~$\zeta$'s.

\begin{figure}
		\begin{center}
		\includegraphics[scale=1]{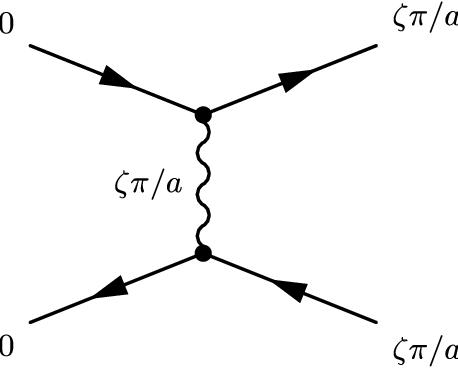}
	\end{center}
	\caption{The leading tree-level taste-exchange interaction, which 
involves the exchange of a gluon with momentum $\zeta\pi/a$ where each 
$\zeta_\mu$ is~0 or~1 but $\zeta^2\ne0$.}
	\label{treelevel-fig}
\end{figure}

As a consequence of the doubling symmetry, the standard low-energy
mode and the fifteen doubler modes must be interpreted as sixteen
equivalent flavors or ``tastes'' of quark. (The sixteen tastes are 
reduced to four
by staggering the quark field; see 
Appendix~\ref{about-staggered-quarks}.)
This unusual
implementation of quark tastes has surprising consequences. Most
striking is that a low-energy quark that absorbs momentum close to
$\zeta\,\pi/a$, for one of the fifteen~$\zeta$'s, is not driven far off
energy-shell.
Rather it is turned into a low-energy quark of another
taste. Thus the simplest process by which a quark changes taste is
the emission of a single gluon with momentum~$q\approx\zeta\,\pi/a$.
This gluon is highly virtual, and therefore
it must immediately be reabsorbed by another quark, whose taste will
also change (see Fig.~\ref{treelevel-fig}).
Taste changes necessarily involve highly virtual gluons,
and so are both suppressed (by $a^2$) and perturbative for typical 
lattice spacings.
One-gluon exchange, with gluon momentum $q\approx\zeta\,\pi/a$, is
the dominant flavor-changing interaction since it is lowest order
in~$\alpha_s(\zeta\pi/a)$ and involves only 4~external quark lines. 
(Processes with more quark lines are suppressed by additional powers of 
$(ap)^3$ where $p$ is a typical external momentum.)
This observation is crucial when trying to
improve naive quarks by removing finite-$a$ errors, as we discuss below 
(see also \cite{lepage1-paper,sinclair-paper,lepage2-paper}).

Sixteen tastes of quark from a single quark field is fifteen too many. 
In the absence of taste exchange, factors of $1/16$ are easily inserted 
into simulations to remove the extra copies. In particular, quark 
vacuum polarization is corrected by replacing the quark determinant in 
the path integral by its $1/16$~root,
\begin{equation} \label{sixteenth-rule}
	\det(\gamma\cdot\Delta +m_0) \to \det(\gamma\cdot\Delta +m_0)^{1/16},
\end{equation}
or, equivalently, by multiplying the contribution from each quark loop 
by~$1/16$. Recent empirical studies of the spectrum of 
$\gamma\cdot\Delta$, and improved versions of it, show how eigenvalues 
cluster into increasingly degenerate multiplets of sixteen  as the 
lattice spacing vanishes, which is necessary if this procedure is to be 
useful for quark vacuum polarization\,\cite{Dslash-spectrum}. The 
treatment of valence quarks is illustrated in 
Appendix~\ref{app-meson-prop}.

\begin{figure*}
	\begin{center}
		\includegraphics[scale=1]{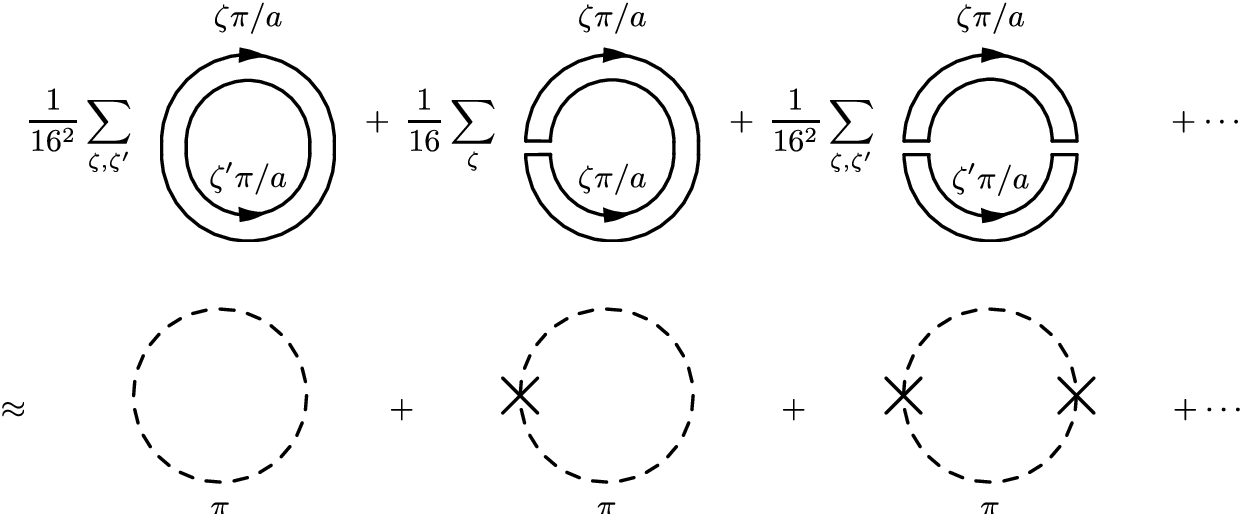}
	\end{center}
	\caption{Infrared contributions from (naive lattice)
	quark vacuum polarization corresponding
	to a pion vacuum polarization loop in a simulation of one flavor.
	The sums are over all infrared sectors, where
	quarks have momenta near $p_\mu=\zeta_\mu\pi/a$ for one of the sixteen 
$\zeta$s
	consisting of just~0s and~1s. The factors of $1/16$ are from the 
$1/16$~rule
	(\eq{sixteenth-rule}) and cancel the sums. Gluons, to all orders, are
	implicit in these diagrams.}
	\label{piloop-fig}
\end{figure*}

It is easy to see how the $1/16$~root corrects for taste in particular 
situations. For example, consider the vacuum polarization of a single 
pion in a theory of only one massless quark flavor. Ignoring taste 
exchange, the dominant infrared contributions come from the quark 
diagrams shown in Fig.~\ref{piloop-fig} (gluons, to all orders, are 
implicit in these diagrams). The first diagram, with its double quark 
loop, has contributions from $16^2$~identical massless pions, 
corresponding to configurations where the quark and antiquark each 
carries a momentum close to an integer multiple of $\pi/a$ (and is 
therefore close to being on shell). This diagram is multiplied by 
$1/16^2$ since there are two quark loops, thereby giving the 
contribution of a single massless pion. The second diagram involves 
quark annihilation into gluons, and has contributions from 
$16$~identical massless pions, corresponding to configurations where 
the quark and antiquark each carries a momentum close to the same 
$\zeta\pi/a$ (so they can annihilate into low-momentum gluons). This 
diagram is multiplied by $1/16$ since there is only one quark loop, 
thereby giving the contribution again of a single massless pion but now 
with an insertion from the gluon decay. Inserting additional 
annihilation kernels results in a geometric series of insertions in the 
propagator of a single massless pion that shifts the pion's mass away 
from zero, as expected in a $U(1)$~flavor theory.

Such patterns are disturbed by taste-exchange interactions like that in 
Fig.~\ref{treelevel-fig}. For example, these interactions cause mixing 
between the different tastes of pion in Fig.~\ref{piloop-fig}. This 
mixing lifts the degeneracy in the pion masses so that different tastes 
of pion are only approximately equivalent. Consequently the 
``$1/16$-root rule'' gives results that correspond only approximately 
to a single pion. Taste-changing interactions are suppressed by~$a^2$, 
however, and detailed analyses, using chiral perturbation theory, 
demonstrate that the impact on physical quantities is therefore also 
suppressed by~$a^2$\,\cite{aubin}. Consequently, while taste exchange 
can lead to such anomalies as unitarity violations, these are 
suppressed in the continuum limit. They are further, and more 
efficiently, suppressed through Symanzik improvement of the action, as 
we now discuss.

\subsection{Tree-Level Symanzik Improvement}
The discretization errors in the naive-quark action come from two 
sources. The more conventional of these corrects the finite-difference 
approximation to the derivatives in the action: one 
replaces\,\cite{naik-paper}
\begin{equation}
	\Delta_\mu  \to  \Delta_\mu - \frac{a^2}{6} \Delta_\mu^3
	\label{naik-term}
\end{equation}
in the naive-quark action (\eq{naive-quark-action}). The 
$a^2$~correction is often referred to as the ``Naik term.''

Less conventional is a correction to remove leading order 
taste-exchange 
interactions~\cite{lepage1-paper,sinclair-paper,lepage2-paper}. As 
discussed in the previous section, these interactions result from the 
exchange of single gluons carrying momenta close to $\zeta\pi/a$ for 
one of the fifteen non-zero $\zeta$s ($\zeta_\mu\in\mathbb{Z}_2$). 
Since these gluons are highly virtual, such interactions are 
effectively the same as four-quark contact interactions and could be 
canceled by adding four-quark operators to the quark action. These 
operators affect physical results in~$\order(a^2)$ since they have 
dimension six. A simpler alternative to four-quark operators is to 
modify the gluon-quark vertex, $\psib \gamma_\mu U_\mu \psi + \cdots$, 
in the original action by introducing a form factor~$f_\mu(q)$ that 
vanishes for (taste-changing) gluons with momenta $q=\zeta\pi/a$ for 
each of the fifteen nonzero~$\zeta$s. In fact the form factor for 
direction~$\mu$ need not vanish when $\zeta_\mu=1$ since the original 
interaction already vanishes in that case. Consequently we want a form 
factor where
\begin{equation}
	f_\mu(q) \to
	\begin{cases}
	1 & \text{for $q\to0$} \\
	0 & \text{for $q\to\zeta\pi/a$ where $\zeta^2\ne0$, $\zeta_\mu=0$}.
	\end{cases}
\end{equation}

We can introduce such a form factor by replacing the link operator 
$U_\mu(x)$ in the action with $\F_\mu U_\mu(x)$
where smearing operator $\F_\mu$ is defined by
\begin{equation}
	\F_\mu \equiv \prod_{\rho\ne\mu}
	\left.\left(
		1 + \frac{a^2\delta^{(2)}_\rho}{4}
	\right)\right|_\text{symm.}
\end{equation}
and $\delta^{(2)}_\rho$ approximates a covariant second derivative when 
acting on link fields:
\begin{align}
\delta^{(2)}_\rho\,U_\mu(x) &\equiv
\frac{1}{a^2} \left(
U_\rho(x)\,U_\mu(x+a\hat\rho)\,U_\rho^\dagger(x+a\hat\mu) \right. \nl
& - 2U_\mu(x)   \nl
& + \left.U_\rho^\dagger(x-a\hat\rho)\,U_\mu(x-a\hat\rho)
\,U_\rho(x-a\hat\rho+a\hat\mu)\right).
\end{align}
This works because $\delta^{(2)}_\rho\approx -4/a^2$ (and $\F_\mu$ 
vanishes) when acting on a link field that carries momentum 
$q_\rho\approx\pi/a$. This kind of link smearing is referred to as 
``Fat7'' smearing in~\cite{doug-papers}.

Smearing the links with $\F_\mu$ removes the leading 
$\order(a^2)$~taste-exchange interactions, but introduces new 
$\order(a^2)$~errors. These can be removed by replacing $\F_\mu$ 
with\,\cite{lepage2-paper}
\begin{equation} \label{Fasqtad}
	\Ft_\mu \equiv \F_\mu -
	\sum_{\rho\ne\mu} \frac{a^2(\delta_\rho)^2}{4}
\end{equation}
where $\delta_\rho$ approximates a covariant first derivative:
\begin{align}
\delta_\rho\,U_\mu(x) &\equiv
\frac{1}{a} \left(
U_\rho(x)\,U_\mu(x+a\hat\rho)\,U_\rho^\dagger(x+a\hat\mu) \right. \nl
& - \left.U_\rho^\dagger(x-a\hat\rho)\,U_\mu(x-a\hat\rho)
\,U_\rho(x-a\hat\rho+a\hat\mu)\right).
\end{align}
The new term has no effect on taste exchange but (obviously) cancels 
the $\order(a^2)$~part of~${\cal F}_\mu$. Correcting the derivative, as 
in \eq{naik-term}, and replacing links by $a^2$-accurate smeared links 
removes all tree-level $\order(a^2)$~errors in the naive-quark action. 
The result is the widely used ``ASQTAD'' action\,\cite{lepage2-paper},
\be
\sum_{x}\, \psib(x)\,
\left(\sum_\mu\gamma_\mu\left(\Delta_\mu(V) - 
\frac{a^2}{6}\Delta_\mu^3(U)\right) +m_0\right)\,\psi(x),
\ee
where, in the first difference operator,
\be
V_\mu(x) \equiv \Ft_\mu \, U_\mu(x).
\ee
In practice, operator $V_\mu$ is usually tadpole 
improved~\cite{tadpole}; in fact, however, tadpole improvement is not 
needed when links are smeared and 
reunitarized~\cite{tsukuba,schladming}.

\subsection{$c$ Quarks}\label{c-errors0}
The tree-level discretization errors in the ASQTAD action 
are~$\order((ap_\mu)^4)$, which are negligible ($<1$\%) for light 
quarks. When applied to $c$~quarks, however, these errors will be 
larger. The most important errors in this case are associated with the 
quark's energy, rather than its 3-momentum, since $c$~quarks are 
typically nonrelativistic in $\psi$s, $D$s and other systems of current 
interest. Consequently $E\approx m \gg\pv$ and the largest errors 
are~$\order((am)^4)$ or 6\% at current lattice spacings 
($am_c\approx0.5$). Such errors show up, for example, in the tree-level 
dispersion relation of the quark where:
\bea
	c^2(0) &\equiv& \lim_{\pv\to 0} \frac{E^2(\pv)-m^2}{\pv^2}  \nonumber 
\\
	&=& 1 + \frac{9}{20}\,(am)^4
	+ \frac{1}{7}\,(am)^6 + \cdots. \label{c2-eq}
\eea

This dominant (tree-level) error can be removed by retuning the 
coefficient of the $a^2\gamma_\mu \Delta_\mu^3$ (Naik) term in the 
action:
\be
\sum_{x}\, \psib(x)
\left(\sum_\mu\gamma_\mu\left(\Delta_\mu -
\frac{a^2}{6}(1+\epsilon)\Delta_\mu^3\right) +m_0\right) \psi(x),
\label{eps-def}
\ee
where parameter $\epsilon$ has expansion
\be
\begin{split}
\epsilon = &-\frac{27}{40}\,(am)^2 + \frac{327}{1120}\,(am)^4
	-\frac{5843}{53760}\,(am)^6 \\
	&+ \frac{153607}{3942400}\,(am)^8 -\cdots.
\end{split}
\label{tuneepsilon}
\ee
With this choice, $c^2=1+\order((am)^{12})$.
Only the first term in $\epsilon$'s expansion is essential for removing 
$(am)^4$~errors; the remaining terms remove errors to higher orders in 
$am$ but leading order in~$v/c$, the quark's typical velocity. In 
practice, terms beyond the first one or two are negligible though 
trivial to include.

Note that the bare mass~$m_0$ in the quark action is related to the 
tree-level quark mass by
\be
m_0 = m\left(1 +\frac{3}{80}\,(am)^4 -\frac{23}{2240}\,(am)^6 + 
\cdots\right)
\ee
for this choice of~$\epsilon$. This formula is useful for extracting 
the corrected bare quark mass,~$m$, from a tuned value of~$m_0$.

To examine the tree-level errors in this modified ASQTAD action, it is 
useful to make a nonrelativistic expansion since, as we indicated, 
heavy quarks are generally nonrelativistic in the systems of interest. 
This expansion is easy because the action is identical in form to the 
continuum Dirac equation, and consequently has the same nonrelativistic 
expansion but with
\be
D_\mu \to \Delta_\mu - \frac{a^2}{6}\,(1+\epsilon)\Delta_\mu^3.
\ee
At tree level,
\be
	\Delta_\mu \to \sinh(aD_t - am)/a,
\ee
when we replace
\be
	\psi\to\exp(-mt)\,\psi,
\ee
where $m$ is the quark mass. This redefinition removes the rest mass 
from all energies. Expanding in powers of $1/m$, we then obtain the 
tree-level nonrelativistic expansion of the modified ASQTAD action:
\begin{equation}
	\begin{split}
		&\sum\psi^\dagger_\mathrm{NR} \left(
		 D_t  - \frac{\Dv^2}{2m}
		 - c_K\,\frac{\Dv^4}{8m^3}
		 -\frac{g\sigma\cdot\Bv}{2m} \right.
		\\
		 &\left. + c_E\,\frac{ig[\Dv\cdot,\Ev]}{8m^2}
		 - c_E\,\frac{g\sigma\cdot(\Dv\times\Ev-\Ev\times\Dv)}{8m^2} + \cdots
		\right) \psi_\mathrm{NR}
		\end{split}
\end{equation}
where
\begin{align}
	c_K &= 1-\frac{9}{10}\,(am)^4 + \frac{29}{280}\,(am)^6-\cdots \\
	c_E &= 1-\frac{3}{80}\,(am)^4 + \frac{23}{2240}\,(am)^6-\cdots.
\end{align}
The coefficients of all terms through order~$1/m$ have no $am$~errors 
at tree level. In particular the $\sigma\cdot\Bv$ term is correct at 
tree level once $m$~is tuned (nonperturbatively). The coefficients of 
the remaining terms shown are off by less than~6\% and~1\% for charm 
quarks on current lattices ($am\approx 1/2$), which should lead to 
errors of order $(am)^4(v/c)^2\approx2$\% or less in charmonium 
analyses and less than~0.5\% for $D$~physics. (Errors should actually 
be considerably smaller than this because of dimensionless factors like 
the~$1/8$ in the $\Dv^4$ term.) Other tree-level errors, which are 
order~$(a\pv)^4$ and~$\epsilon(a\pv)^2/6$, should also be less than~1\% 
for~$a\approx0.1$\,fm.

\section{Symanzik Improvement at One-Loop Order}
One-loop errors, which are $\order(\alpha_s (ap)^2)$, should be no more 
than a percent or so for light quarks and current lattice spacings. 
Taste-changing errors, however, have generally been much larger than 
expected. These errors also bear directly on the validity of the 
$1/16$-root trick for accounting for taste in the vacuum polarization. 
For this reason it is worth trying to further suppress taste-exchange 
errors beyond what has been accomplished at tree level. We show how to 
do this in this section.

The only other one-loop effects that are important at the level of a 
few percent are $\alpha_s(am)^2$ for heavy quarks, and, in particular, 
the charm quark. We show how to correct for these errors in the 
Section~\ref{c-errors1}.

\subsection{Light Quarks}
\begin{figure*}[t]
		\begin{center}
		\includegraphics[scale=1]{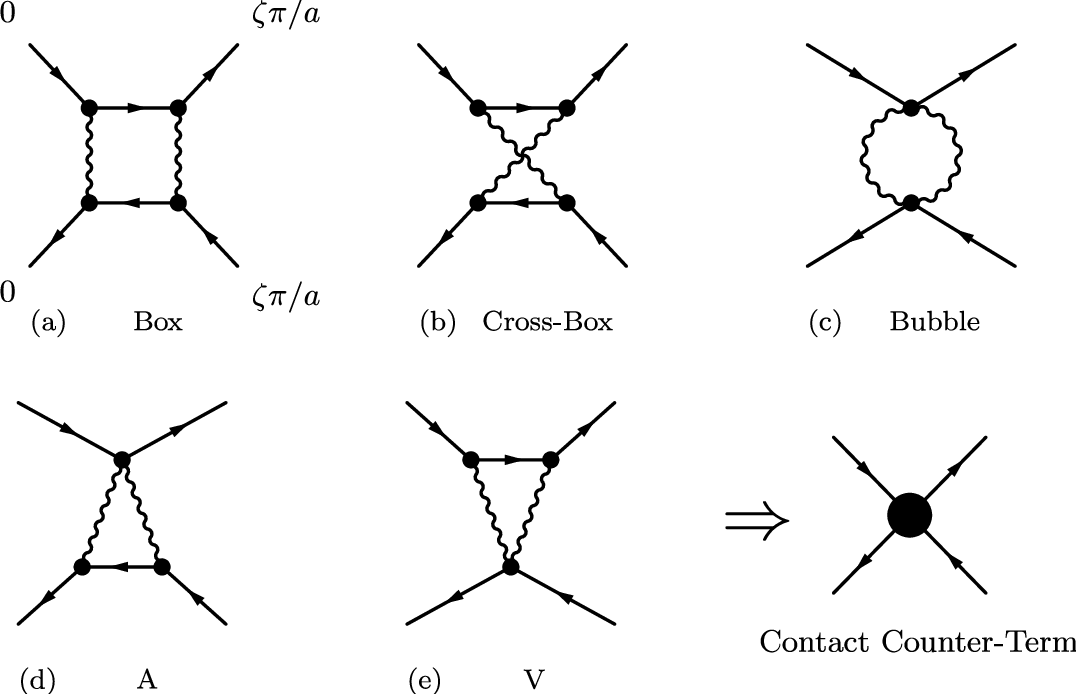}
	\end{center}
	\caption{The only one-loop diagrams that result in
	$q\overline{q}\to q\overline{q}$ taste exchange for ASQTAD quarks.
	Diagrams with additional external quark lines are suppressed by
	additional powers of~$a^2$.}
	\label{oneloop-fig}
\end{figure*}

One-loop taste exchange comes only from the diagrams shown in 
Fig.~\ref{oneloop-fig}; other one-loop diagrams vanish because of the 
tree-level improvements. Again the gluons in these diagrams transfer 
momenta of order~$\zeta\pi/a$, and so are highly virtual. Consequently, 
the correction terms in the corrected action that cancel these will 
involve current-current interactions that conserve overall taste 
(because of momentum conservation). There are 28 such terms in the 
massless limit for the staggered-quark action (see 
Appendix~\ref{app-oneloop}):
\begin{align}
&\begin{array}{r@{\,}l@{}c@{\,}l}
2\,\Delta\mathcal{L}_\text{contact} =
& d_{5}^{( 5\mu)} \left|\mathcal{J}_{5}^{( 5\mu)}\right|^2
 &+& d_{5\mu\nu}^{( 5\nu)} \left|\mathcal{J}_{5\mu\nu}^{( 
5\nu)}\right|^2\\
+& d_{\nu}^{(\mu\nu)} \left|\mathcal{J}_{\nu}^{(\mu\nu)}\right|^2
 &+& d_{5\nu}^{( 5\mu\nu)} \left|\mathcal{J}_{5\nu}^{( 
5\mu\nu)}\right|^2\\
+& d_{\mu\nu}^{(\nu)} \left|\mathcal{J}_{\mu\nu}^{(\nu)}\right|^2
 &+& d_{1}^{(\mu)} \left|\mathcal{J}_{1}^{(\mu)}\right|^2\\
+& d_{5\mu}^{( 5)} \left|\mathcal{J}_{5\mu}^{( 5)}\right|^2
 &+&  \\
+& d_{1}^{( 5\mu)} \left|\mathcal{J}_{1}^{( 5\mu)}\right|^2
 &+& d_{\mu\nu}^{( 5\nu)} \left|\mathcal{J}_{\mu\nu}^{( 
5\nu)}\right|^2\\
+& d_{\nu}^{( 5\mu\nu)} \left|\mathcal{J}_{\nu}^{( 5\mu\nu)}\right|^2
 &+& d_{5\nu}^{( \mu\nu)} \left|\mathcal{J}_{5\nu}^{( 
\mu\nu)}\right|^2\\
+& d_{5\mu\nu}^{(\nu)} \left|\mathcal{J}_{5\mu\nu}^{(\nu)}\right|^2
 &+& d_{5}^{(\mu)} \left|\mathcal{J}_{5}^{(\mu)}\right|^2\\
&
 &+& d_{\mu}^{( 5)} \left|\mathcal{J}_{\mu}^{( 5)}\right|^2\\
\end{array}\label{e:tastechange}\\
&\phantom{\Delta\mathcal{L}\,\,}+\left(\text{color octet $\mathcal{J}$s 
with $d_n^{(t)}\to \tilde{d}^{(t)}_n$}\right),\notag
\end{align}
where sums over indices are implicit and $\nu\ne\mu$. The currents 
$\mathcal{J}^{(s)}_n$ in the first 14 operators are color-singlets with 
taste~$s$ and spinor structure~$n$, while the last 14~operators are the 
same but with color octet currents. The precise definitions of these 
currents are given in Appendix~\ref{app-stagg-currents}.

\begin{table}
	\begin{tabular}{lcccc} \hline\hline
	& \multicolumn{3}{c}{Unimproved Gluons} & Improved Gluons \\
	 & ASQTAD & HISQ & HYP & ASQTAD \\ \hline
	Octet: \\
	\,\,\,\,\,$\tilde{d}_{5}^{( 5\mu)}$ & 1.41 & 0.19 & 0.02 & 0.77\\
	\,\,\,\,\,$\tilde{d}_{5\mu\nu}^{( 5\nu)}$ & 0.46 & 0.00 & 0.08 & 
0.25\\
	\,\,\,\,\,$\tilde{d}_{1}^{( 5\mu)}$ & 0.23 & 0.01 & 0.00 & 0.16\\
	\,\,\,\,\,$\tilde{d}_{\mu\nu}^{( 5\nu)}$ & 0.38 & 0.02 & 0.01 & 0.27\\
	\,\,\,\,\,$\tilde{d}_{\nu}^{(\mu\nu)}$ & 0.34 & 0.08 & 0.04 & 0.19\\
	\,\,\,\,\,$\tilde{d}_{5\nu}^{( 5\mu\nu)}$ & 0.35 & 0.03 & 0.02 & 
0.19\\
	\,\,\,\,\,$\tilde{d}_{\nu}^{( 5\mu\nu)}$ & 0.08 & 0.00 & 0.01 & 0.05\\
	\,\,\,\,\,$\tilde{d}_{5\nu}^{( \mu\nu)}$ & 0.20 & 0.00 & 0.01 & 0.13\\
	\,\,\,\,\,$\tilde{d}_{\mu\nu}^{(\nu)}$ & 0.20 & 0.01 & 0.02 & 0.11\\
	\,\,\,\,\,$\tilde{d}_{1}^{(\mu)}$ & 0.31 & 0.01 & 0.01 & 0.17\\
	\,\,\,\,\,$\tilde{d}_{5\mu\nu}^{(\nu)}$ & 0.06 & 0.00 & 0.00 & 0.04\\
	\,\,\,\,\,$\tilde{d}_{5}^{(\mu)}$ & 0.07 & 0.00 & 0.01 & 0.04\\
	\,\,\,\,\,$\tilde{d}_{5\mu}^{( 5)}$ & 0.17 & 0.00 & 0.01 & 0.09\\
	\,\,\,\,\,$\tilde{d}_{\mu}^{( 5)}$ & 0.06 & 0.00 & 0.00 & 0.04\\
	\\ Singlet: \\
	\,\,\,\,\,$d_{5}^{( 5\mu)}$ & 0.76 & 0.10 & 0.01 & 0.41\\
	\,\,\,\,\,$d_{1}^{( 5\mu)}$ & 0.12 & 0.00 & 0.00 & 0.09\\
	\,\,\,\,\,$d_{\nu}^{(\mu\nu)}$ & 0.18 & 0.05 & 0.02 & 0.10\\
	\,\,\,\,\,$d_{\nu}^{( 5\mu\nu)}$ & 0.04 & 0.00 & 0.00 & 0.03\\
	\,\,\,\,\,$d_{\mu\nu}^{(\nu)}$ & 0.11 & 0.01 & 0.01 & 0.06\\
	\,\,\,\,\,$d_{5\mu\nu}^{(\nu)}$ & 0.03 & 0.00 & 0.00 & 0.02\\
	\,\,\,\,\,$d_{5\mu}^{( 5)}$ & 0.09 & 0.00 & 0.00 & 0.05\\
	\\ Avg $d$,$\tilde{d}$ & 0.23 & 0.02 & 0.02 & 0.13\\
	$a^2\Delta m_\pi^2$ & 0.023(1) & 0.011(1) & 0.009(1)\\
	\hline\hline \end{tabular}
\caption{$\order(\alpha_s)$ coefficients for the couplings in 
$\lag_\mathrm{contact}$ (\eq{e:tastechange}) for three different 
light-quark actions and massless quarks. Terms not listed have zero 
coefficient in this order. Results are given for the unimproved 
(Wilson) gluon action, and, in the ASQTAD case, also for the standard 
L\"uscher-Weisz action. Results for the average coefficient are also 
given, as are the splittings $\Delta m_\pi^2$ between the 3-link pion 
and the Goldstone pion. The pion masses are from simulations with 
uncorrected gluons at $1/a\approx 1/10$\,fm with no vacuum 
polarization.}
\label{d-table}
\end{table}

We have computed the coupling constants~$d_n^{(s)}$ 
and~$\tilde{d}_n^{(s)}$ for the tree-level improved theory; the results 
are presented in Table~\ref{d-table} (see also 
Appendix~\ref{app-oneloop}). Armed with these one could incorporate 
$\Delta\mathcal{L}_\text{contact}$ into simulations, removing all 
one-loop taste exchange. This procedure is complicated to implement, 
however. A simpler procedure is suggested by the results 
in~\cite{hyp-paper}. That analysis shows that repeated smearing of the 
links further reduces the mass splittings between pions of different 
taste.

This result is not surprising given the perturbative origins of taste 
exchange. The dominant taste-exchange interactions (in ASQTAD) come 
from the one-loop diagrams in Fig.~\ref{oneloop-fig}. The largest 
contributions to these diagrams come from large loop momenta, 
$k=\order(\zeta\pi/2a)$. Smearing the links introduces form factors 
that suppress high momenta, leaving low momenta unchanged, and 
therefore should suppress this sort of one-loop correction. The links 
in the ASQTAD action are smeared, but the operator $\Ft_\mu$, which 
smears links in the $\mu$~direction, introduces additional links in 
orthogonal directions (to preserve gauge invariance) that are not 
smeared. Smearing the links multiple times guarantees that all links 
are smeared, and high momenta are suppressed.

There are three problems associated with multiple smearing. One is that 
it introduces new $\order(a^2)$ discretization errors, and these errors 
grow with additional smearing. The action presented in~\cite{hyp-paper} 
suffers from this problem. The $a^2$~errors can be avoided by using an 
$a^2$-improved smearing operator, such as $\Ft_\mu$ (\eq{Fasqtad}). 
Replacing smearing operator
\begin{equation}
	\Ft_\mu \to \Ft_\mu\, \Ft_\mu
\end{equation}
in the quark action, for example, smears all links but does not 
introduce new tree-level $a^2$~errors.

The second problem with multiple smearings is that they replace a 
single link in the naive action by a sum of a very large number of 
products of links. This explosion in the number of terms does not 
affect single-gluon vertices on quark lines, by design, but leads to a 
$\sqrt{N}$-growth in the size of two-gluon vertices where $N$~is the 
number of terms in the sum. The $\sqrt{N}$-growth enhances one-loop 
diagrams with two-gluon vertices, canceling much of the benefit 
obtained from smearing.
This problem is remedied by reunitarizing the smeared link operator: 
that is we replace
\begin{equation}
	\Ft_\mu \to \Ft_\mu \,{\cal U} \,\Ft_\mu
\end{equation}
where operator $\cal U$ unitarizes whatever it acts on. A smeared link 
that has been unitarized is bounded (by unity) and so cannot suffer 
from the $\sqrt{N}$ problem. Although we verified  that it is 
unnecessary to make the unitarized link into an 
$SU(3)$~matrix\,---\,simple unitarization is adequate\,---\,the results 
given in this paper use links that are projected back onto~$SU(3)$. 
Reunitarizing has no effect on the single-gluon vertex, and therefore 
does not introduce new $\order(a^2)$~errors.

The doubly smeared operator is simplified if we rearrange it as follows
\begin{equation} \label{Fhisq}
	\Ftt_\mu \equiv \left(\F_\mu - \sum_{\rho\ne\mu}
		\frac{a^2(\delta_\rho)^2}{2}\right)\,{\cal U}\,\F_\mu,
\end{equation}
where the entire correction for $a^2$~errors is moved to the outermost 
smearing. Our new ``HISQ'' discretization of the quark action is 
therefore
\be \label{hisq-def}
\sum_{x}\, \psib(x)\,
\left(\gamma\cdot\mathcal{D}^\mathrm{HISQ} +m\right)\,\psi(x),
\ee
where
\be
\mathcal{D}^\mathrm{HISQ}_\mu \equiv
\Delta_\mu(W) - \frac{a^2}{6}(1+\epsilon)\Delta_\mu^3(X)
\ee
and now, in the first difference operator,
\be \label{W-def}
W_\mu(x) \equiv \Ftt_\mu \, U_\mu(x),
\ee
while in the second,
\be \label{X-def}
X_\mu(x) \equiv {\cal U}\,\F_\mu \, U_\mu(x).
\ee

The third problem with smearing is that previous work has only 
demonstrated its impact on the mass splittings between pions with 
different tastes. For small masses, there are only five degenerate 
multiplets of pion. Consequently showing that smearing reduces the 
splittings between the pion multiplets does not guarantee that all 
28~taste-exchange terms in the quark action (\eq{e:tastechange}) are 
suppressed; and, therefore, it does not guarantee that taste-exchange 
effects are generally suppressed (other than in the pion mass 
splittings).

To examine this issue, we recomputed the one-loop coefficients for all 
of the leading taste-exchange correction terms in the action 
(\eq{e:tastechange}) using the HISQ action, as well as the HYP action 
from~\cite{hyp-paper}. The results are derived in 
Appendix~\ref{app-oneloop} and presented in Table~\ref{d-table}. The 
HISQ coefficients are on average an order of magnitude smaller than the 
ASQTAD coefficients. Consequently the residual taste-exchange one-loop 
corrections in the HISQ~action are probably smaller than two-loop 
corrections and  possibly also dimension-8 taste-exchange operators at 
current lattice spacings. The HYP action gives coefficients similar in 
size to~HISQ.

\subsection{$c$ Quarks}\label{c-errors1}
Many new taste-preserving operators would need to be added to the HISQ 
action in order to remove all order~$\alpha_s a^2$ errors. The leading 
operators are listed in Table~\ref{op-table}. None of these is relevant 
at the few percent level for light quarks, but terms that enter in 
relative order $\alpha_s (am_c)^2$ might change answers by as much 
as~5--10\% for $c$~quarks on current lattices, where $am_c\approx0.5$ 
and $\alpha_s\approx 1/3$. High-precision work requires that these 
errors be removed.

Nonrelativistic expansions of the various operators in 
Table~\ref{op-table} show that only the first, the Naik term, can cause 
errors in order $\alpha_s(am_c)^2$. All of the others result in 
harmless renormalizations or are suppressed by additional powers of 
$v/c$, and so contribute only at the level of~2--3\% or less for 
$\psi$s and less than~$1$\% for $D$s. We can remove all $\alpha_s 
(am_c)^2$ errors by including radiative corrections in the 
$\epsilon$~parameter that multiplies the Naik term (\eq{eps-def}).

There are two ways to compute the radiative correction to~$\epsilon$. 
One is nonperturbative: $\epsilon$ is adjusted until the relativistic 
dispersion relation,
\be \label{c2-def}
c^2(\pv) \equiv \frac{E^2(\pv)-m^2}{\pv^2} = 1,
\ee
computed in a meson simulation is valid for all low 
three-momenta~$\pv$.

A simpler procedure is to compute the one-loop correction to $\epsilon$ 
using perturbation theory\,---\,by requiring, for example, the correct 
dispersion relation for a quark in one-loop order. The result would 
have the form
\be
	\epsilon = \epsilon_1 \alpha_s - \frac{27}{40}(am)^2  +
	 				\order(\alpha_s^2,(am)^4)
\ee
where $\epsilon_1$ depends upon~$(am)^2$. As we will show, $\epsilon_1$ 
turns out to be negligibly small for the HISQ action (but not for 
ASQTAD).

\begin{table}
\begin{tabular}{ccccc} \hline\hline
	Operator &\quad\quad& $\psi$ Physics &\quad\quad& $D$ Physics
	\\ \hline
	$\,a^2\,\psib D_\mu^3\gamma_\mu\psi$
	&& $\alpha_s(am)^2$ && $\alpha_s(am)^2$
	\\ \hline
	$ a^2m \psib\sigma\cdot g F\psi$
	&& $\alpha_s(v/c)^2(am)^2$ && $\alpha_s(v/c)(am)^2$
	\\
	$ a^2\psib D^2 D\cdot\gamma\psi$
	\\
	$ a^2\psib (D\cdot\gamma)^3 \psi$
	\\
	$ a^2m \psib D^2 \psi$
	\\
	$ a^2\psib \sigma\cdot gF\,D\cdot\gamma\psi$
	\\ \hline
	$a^2\psib D\cdot gF\cdot \gamma\psi$
	&& $\alpha_s(v/c)^2(am)^2$ && $\alpha_s(v/c)^2(am)^2$
	\\
	$ a^2(\psib\gamma\psi)^2$
	\\
	$ a^2(\psib\gamma\gamma_5\psi)^2$
	\\
	\hline \hline
\end{tabular}
\caption{Relative errors associated with the leading (for charm quarks) 
taste-preserving operators that enter the HISQ action in one-loop 
order. Note that $\alpha_s\approx1/3$, $(am)^2\approx1/4$, and 
$(v/c)^2\approx1/3$ for $\psi$s  ($1/10$ for $D$s) for lattice spacings 
of order~0.1\,fm. The errors listed are relative to the binding energy 
which is of order 500\,MeV for both~$\psi$s and~$D$s. Some of the 
operators are related to others in the same grouping by the equations 
of motion, and so are redundant.}
\label{op-table}
\end{table}

\section{Applications: Bounding Taste Exchange}
We tested our perturbative analysis of the suppression of taste 
exchange by computing the pion mass taste splittings for ASQTAD, HISQ, 
and a few other variations~\cite{quentin-thesis}. Some of these results 
are shown in Table~\ref{d-table} where we list the mass splitting 
$\Delta m_\pi^2$ between the 3-link pions and the 0-link (Goldstone) 
pion from a quenched simulation with unimproved (Wilson) glue at 
$a=1/10$\,fm and quark mass~$am=0.03$~\cite{ukqcd-w}. As expected, 
small mass splittings are correlated with small values for the 
coefficients of the one-loop taste-exchange interactions 
(\eq{e:tastechange}). These results confirm: a) that taste-exchange 
interactions are perturbative in origin; b) that these interactions 
will be suppressed by a factor of order~$\alpha_s\approx1/3$ in the 
HISQ action relative to the ASQTAD action at current lattice spacings; 
and c) the HISQ action will be more accurate than other current 
competitors because it has negligible one-loop taste exchange and no 
tree-level $\order(a^2)$~errors.

We have also compared meson splittings for ASQTAD and HISQ valence 
quarks in full QCD simulations with $n_f=3$~light-quark (ASQTAD) vacuum 
polarization. We find that the pseudoscalar splittings are 3.6(5)~times 
smaller with HISQ than with ASQTAD on lattices with~$a=1/8$\,fm, when 
both valence quarks are $s$~quarks. The splittings are~3.1(7)~times 
smaller on our $1/11$\,fm~lattices.

Any taste-exchange effect will be 3--4~times smaller in HISQ, and so 
measuring the ASQTAD--HISQ difference in any quantity allows us to 
bound taste-exchange errors in each formalism. To illustrate how such 
comparisons are done, we have computed the mass of the $\phi$~meson and 
the pseudovector decay constant~$f_{\eta_s}$ of the $\eta_s$~meson 
using ASQTAD and HISQ valence quarks for each of our two lattice 
spacings.

The $\phi$ is unstable and so its mass in a simulation is sensitive to 
the volume of the lattice (it is \emph{not} ``gold-plated'' in the 
sense of~\cite{ratio-paper}). We have not corrected for this volume 
dependence, but have simulated using a fixed physical volume. 
Consequently results for the mass should extrapolate to the same value 
as $a\to0$, but this value may be a few tens of~MeV larger than the 
physical mass. The $\eta_s$ is an easy-to-analyze substitute for 
a~$\pi$; it is a $0^{-+}$ $s\overline{s}$ meson, but without 
valence-quark annihilation, and therefore hypothetical. We tuned the 
$s$~mass so that the $\eta_s$~masses in each case were all identical 
(696\,MeV) to within~$\pm1$\,MeV (statistical/fitting errors only). We 
evaluated the $\eta_s$~decay constant by computing the matrix element 
of the pseudoscalar density and multiplying by the bare quark mass 
(see~\cite{fk-fpi}, for example).

Our results, shown in Table~\ref{etas-table}, suggest significant 
taste-exchange errors for ASQTAD, particularly in the $1/8$\,fm 
simulation. Looking first at the decay constant~$f_{\eta_s}$, the 
ASQTAD result on the coarse lattice is~6(3)\% larger than the ASQTAD 
result on the fine lattice, where taste-exchange effects should be at 
least a factor of two smaller. The uncertainty in the discrepancy is 
large here because of uncertainties in the values of the lattice 
spacings. Far more compelling is the 4.6(6)\%~discrepancy between 
ASQTAD and HISQ on the coarse lattice, which can be measured far more 
accurately than differences between simulations with different lattice 
spacings. This discrepancy is reduced to 1.9(4)\% on the fine lattice, 
as expected since $a^2$ is half as large. Since errors in HISQ should 
be 3--4~times smaller than in ASQTAD, these discrepancies imply that 
taste-exchange errors in the decay constant could be as large as~5--6\% 
for ASQTAD but only~1--2\% for HISQ when $a=1/8$\,fm. The errors are 
half that size when~$a=1/11$\,fm. Results from both formalisms are 
consistent with an extrapolated value of~180\,(4)\,MeV.

Results for the $\phi$~mass are very similar. These give a final mass 
of 1.05(2)\,GeV, which is 30(20)\,MeV above the mass~1.019\,GeV from 
experiment (and consistent with expectations for finite-volume 
effects). These analyses taken together indicate that HISQ is 
delivering 1--2\%~precision already on the coarse lattice, while a 
significantly smaller lattice spacing (and much more costly simulation) 
is needed to achieve similar precision from ASQTAD. HISQ $a^2$~errors 
should be less than~1\% on the~$1/11$\,fm lattice for light quarks.

Our tests are only partial because ASQTAD quarks were used for the 
vacuum polarization in both the HISQ and ASQTAD analyses, but the 
effects of changes in the vacuum polarization are typically 3--5~times 
smaller than the effects caused by the same changes in the valence 
quarks. The agreement between our HISQ results from different lattice 
spacings indicates that finite-$a$ errors from the vacuum polarization 
are not large for these lattice spacings.

\begin{table}
	\begin{tabular}{ccccc}\hline\hline
		& $a$ (fm) & ASQTAD & HISQ & ASQTAD$/$HISQ \\ \hline
		$f_{\eta_s}$ (MeV) & 1/8 &  196(4) & 187(4) & 1.046\,(6) \\
		                   & 1/11 & 185(3) & 182(3) & 1.019\,(4) \\
		\\
		$m_\phi$ (GeV) & 1/8 & 1.119(24) & 1.076(23) & 1.040(9) \\
		              & 1/11 & 1.057(18) & 1.052(16) & 1.005(6)\\
		\hline \hline
	\end{tabular}
	\caption{Simulation results for decay constant $f_{\eta_s}$ and 
$\phi$~mass using either ASQTAD or HISQ for the valence quarks from 
full QCD simulations (with $n_f=3$ ASQTAD quarks). Lattice spacing 
errors cancel in the ratio ASQTAD to HISQ computed on the same gluon 
configurations; thus errors are much smaller for the ratio.}
	\label{etas-table}
\end{table}

In past work, taste-exchange errors have been estimated by comparing 
results from different lattice spacings, relying upon the fact that 
these errors vanish as $a\to0$. This example illustrates how the same 
errors can be reliably estimated by comparing ASQTAD with HISQ results 
from simulations using only a single lattice spacing. This approach to 
estimating taste-exchange errors is very efficient. It is almost 
certainly the simplest way to quantify uncertainties about taste 
exchange and the validity of the $1/16$-root trick for vacuum 
polarization.

\section{Applications: $c$ Quarks and Charmonium}
As argued in Sections~\ref{c-errors0} and~\ref{c-errors1}, we expect 
the ASQTAD action and, especially, the HISQ action to work well for 
$c$~quarks even though $am_c$ is typically of order 0.4 or larger on 
current lattices. To achieve high precision (few percent or better), we 
must tune the Naik term's renormalization parameter~$\epsilon$ 
(\eq{eps-def}). Here we do this nonperturbatively by computing the 
speed of light squared, $c^2(\pv)$ (\eq{c2-eq}), for the $\eta_c$ in 
simulations with various values of~$\epsilon$ and tuning $\epsilon$ 
until~$c^2=1$. Our results are summarized in Table~\ref{c2-table}.

\begin{table}
	\begin{center}\begin{tabular}{r c c c r @{\,(} r @{)}}
		\hline\hline
		& $\epsilon$ & $am$ & $(\pv/p_\mathrm{min})^2$ &
			\multicolumn{2}{c}{$c^2(\pv)$} \\ \hline
		ASQTAD ($1/11$\,fm):
		& 0 & 0.38 & 2 & 0.962 & 9 \\
		& 0.3 & 0.38 & 2 & 1.020 & 11 \\
		& 0.4 & 0.38 & 2 & 1.036 & 11 \\
		\\
		HISQ  ($1/11$\,fm):
		& 0 & 0.43 & 2 & 1.029 & 11 \\
		& $-0.115$ & 0.43 & 1 & 0.985 & 16 \\
		& $-0.115$ & 0.43 & 2 & 0.992 & 13 \\
		& $-0.115$ & 0.43 & 3 & 1.014 & 15 \\
		& $-0.115$ & 0.43 & 4 & 0.991 & 11 \\
		\\
		HISQ  ($1/8$\,fm):
		& $0$  & 0.67 & 1 & 1.190 & 20 \\
		& $-1$ & 0.67 & 1 & 0.560 & 10 \\
		& $-0.35$ & 0.67 & 1 & 0.904 & 15 \\
		& $-0.28$ & 0.67 & 1 & 0.950 & 15 \\
		& $-0.21$ & 0.66 & 1 & 1.008 & 13 \\
		& $-0.21$ & 0.66 & 2 & 1.017 & 10 \\
		& $-0.21$ & 0.66 & 3 & 1.019 & 12 \\
		& $-0.21$ & 0.66 & 4 & 1.007 & 7 \\
		\hline\hline
	\end{tabular}\end{center}
	\caption{The speed of light squared, $c^2(\pv)$, computed from the 
$\eta_c$'s dispersion relation in simulations using different quark 
actions. Results are given for different values of the Naik term's 
renormalization parameter~$\epsilon$ (\eq{eps-def}), the lattice 
spacings~$a$, and the meson's three-momenta~$\pv$ (in units of the 
smallest momentum on the lattice which was roughly 0.5\,GeV in each 
case).
	}
	\label{c2-table}
\end{table}

\begin{figure}
		\begin{center}
		\includegraphics{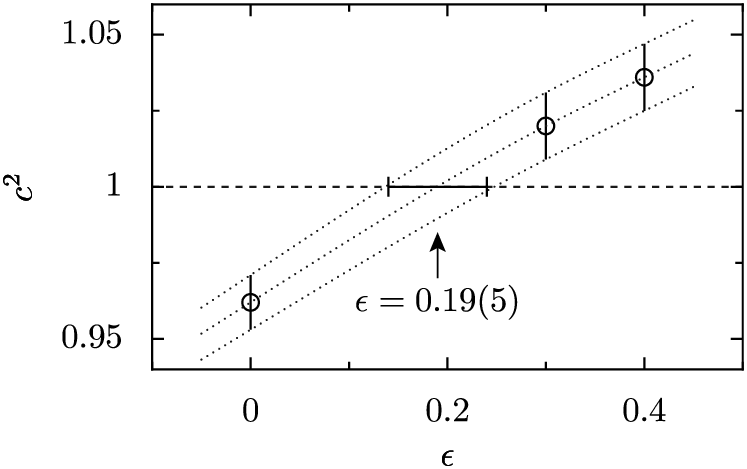}
	\end{center}
	\caption{The speed of light squared, $c^2(\pv)$, computed for an
	$\eta_c$ meson
	using the ASQTAD action with $\pv=(p_\mathrm{min},p_\mathrm{min},0)$ 
and
	different Naik-term renormalizations~$\epsilon$, where 
$p_\mathrm{min}$ is the
	smallest lattice momentum possible (approximately 500\,MeV here). The 
correct
	value, $\epsilon=0.19(5)$, occurs at the intersection of the 
interpolating
	line with the line $c^2=1$, as shown. Here $am_c=.38$.
	}
	\label{c2-eps-fig}
\end{figure}

The ASQTAD results on the fine lattices show only small errors in $c^2$ 
even with $\epsilon=0$. Values for different $\epsilon$s are plotted in 
Fig.~\ref{c2-eps-fig}, together with an interpolating curve. These data 
indicate that the optimal choice  is $\epsilon=0.19(5)$ for this 
lattice spacing and mass. The tree-level prediction for $\epsilon$, 
from \eq{tuneepsilon}, is $-0.10$, which indicates that radiative 
corrections in $\epsilon$ are of order $1\times\alpha_s\approx0.3$, as 
expected.

The HISQ results show even smaller $c^2$~errors on the fine lattices. 
Tuning to $\epsilon=-0.1155$, the tree-level value given 
by~\eq{tuneepsilon} for $am_c=0.43$, removes all errors in~$c^2$ at the 
level of~1\%. This suggests that one-loop and higher-order radiative 
corrections in $\epsilon$ are negligible for HISQ compared with the 
tree-level corrections. The dominance of tree-level contributions is 
confirmed by our HISQ analysis using the coarser lattice spacing, where 
$am_c\approx0.66$. Here, we found the optimal value $\epsilon=-0.22(3)$ 
again by tuning~$\epsilon$ until $c^2=1$ for a low-momentum~$\eta_c$. 
(We overshot slightly and did simulations for $\epsilon=-0.21$ rather 
than~$-0.22$.) Our tuned~$\epsilon$ compares quite well with the 
tree-level prediction of~$-0.246$ from \eq{tuneepsilon}. Consequently 
it is quite likely that the tree-level formula is sufficiently accurate 
for most practical applications today.

Note that setting $\epsilon=-1$ cancels out the Naik term completely. 
Tree-level errors are then order~$a^2$ rather than order~$a^4$, as in 
the HISQ action. Table~\ref{c2-table} shows that these $a^2$~errors 
cause $c^2$ to be off by almost a factor of~2. This example underscores 
the importance of using $a^2$-improved actions in high-precision work.

\begin{figure}
		\begin{center}
		\includegraphics[scale=1]{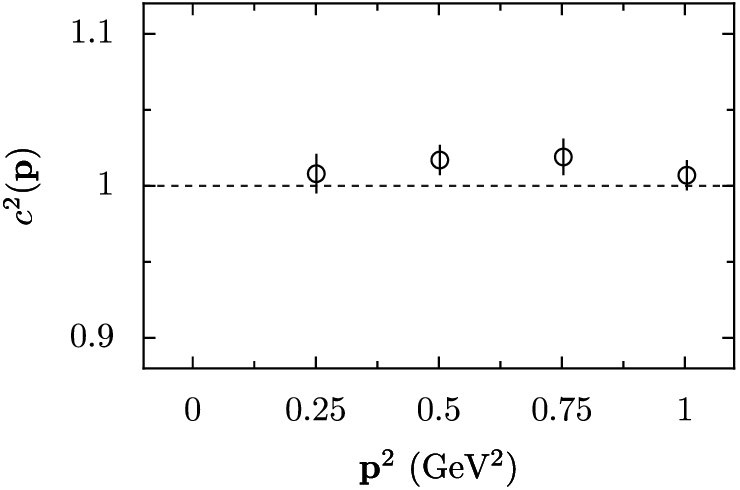}
	\end{center}
	\caption{The speed of light squared, $c^2(\pv^2)$, for $\eta_c$ mesons 
at different momenta on a lattice where $am_c\approx0.66$ for the HISQ 
action with $\epsilon= -0.21$. A comparison with other results from 
Table~\ref{c2-table} suggests that choosing $\epsilon=-0.22$ would move 
all points down onto the line $c^2=1$ to within errors.}
\label{c2-p2-fig}
\end{figure}

On the coarse lattice, we use the $c^2(\pv)$ with $p=p_\mathrm{min}$, 
the smallest nonzero momentum on our lattice, to tune~$\epsilon$. It is 
important to verify that the tuned action gives the correct dispersion 
relation for other momenta as well. The data in Fig.~\ref{c2-p2-fig} 
demonstrate that errors are less than a couple percent for meson 
momenta out to~1\,GeV even on the coarse lattice where $am_c=0.66$. 
Indeed, the errors would probably have been smaller ($<1$\%) had we 
tuned~$\epsilon$ a little more accurately.

\begin{figure}
	\begin{center}
		\includegraphics[scale=1]{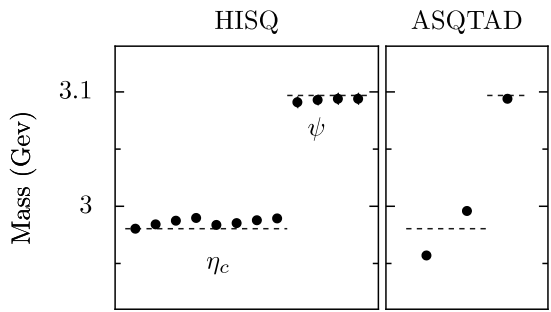}
	\end{center}
	\caption{Masses for different tastes of the~$\eta_c$ and~$\psi$ using 
HISQ and ASQTAD $c$~quarks with $\epsilon=0$ on a $1/11$\,fm lattice 
($am_c\approx0.4$). For HISQ, $\eta_c$s are given (from left to right) 
for the 0, 1, 2, and 3~spatial link tastes, without and then with a 
temporal link; only the spatial splittings are given for the~$\psi$. 
For ASQTAD, $\eta_c$ results are given for only the 0-link and 
1-temporal-link operators, and $\psi$ results  for only the 0-link 
operator. The dashed lines indicate the results from experiment. Error 
bars are of order the size of the plot symbols.}
	\label{hfs-fig}
\end{figure}

We examined the spectrum of the $\psi$~meson family to further test the 
precision of our formalism. We used the $1/11$\,fm lattices where 
$am_c\approx0.4$, and therefore we expect errors of order 1--2\% of the 
binding energy, or~5--10\,MeV (see Section~\ref{c-errors1}). A 
particularly sensitive test is the hyperfine mass splitting between 
the~$\psi$ and~$\eta_c$. We tuned the bare $c$~mass until the mass of 
the local (and lightest)~$\eta_c$ in our simulation agreed with 
experiment. Since there are no other free parameters in our action, the 
simulation then predicts a mass for the~$\psi$. As is clear from 
Fig.~\ref{hfs-fig}, the $\psi$~mass in the simulation is quite 
accurate\,---\,certainly within the 5--10\,MeV we expected.

The data shown in this last figure also bound taste-exchange errors. As 
shown there, different tastes of the $\eta_c$ have different masses, 
because of taste-exchange interactions. The maximum spread in the 
masses for the HISQ action, however, is only 9\,MeV. This is much 
smaller than the spread from the ASQTAD action, where the splitting 
between, for example, the 0-link and 1-temporal-link pion masses is 
40\,MeV (compared to only 3\,MeV for the same splitting with HISQ). 
(Uncorrected staggered quarks also show very large 
splittings~\cite{japan-psi}.) Note also that the spread in the~$\psi$s 
is three times smaller than the spread in the~$\eta_c$s. This is 
typical; the masses of mesons other than pseudoscalars are much less 
sensitive to taste-exchange effects.

The $\psi-\eta_c$ hyperfine splitting, for 0-link mesons in each case, 
is~109(3)\,MeV in our $1/11$\,fm simulation with tuned~$\epsilon$ 
($=-0.115$). The 3\%~uncertainty is almost entirely due to tuning 
uncertainties in the lattice spacings since these uncertainties enter 
twice: once for converting the splitting from lattice to physical 
units, and once through uncertainties in the $c$~mass, which are 
themselves controlled by uncertainties in~$a^{-1}$~\cite{nrqcd}.
Our HISQ result is somewhat smaller than the current experimental 
result of~117(1)\,MeV~\cite{etac-data}, but it needs three further 
corrections. (It is also worth noting that the current Particle Data 
Group average of~117\,MeV is somewhat larger than the most recent 
experiments which find values in the range 113--115\,MeV with 
uncertainties of~1--2\,MeV; see~\cite{etac-data}.)

The first correction comes from the operators in Table~\ref{op-table} 
that must be added to the HISQ lagrangian in order to remove further 
discretization errors. Of these the most important for the hyperfine 
splitting is
\be \label{L-hfs}
\delta\lag_\mathrm{hfs} = c_\mathrm{hfs} \,a^2 m_c \,\psib \sigma\cdot 
F \psi
\ee
where $c_\mathrm{hfs}=\order(\alpha_s)$. This will affect the hyperfine 
splitting in relative order~$\alpha_s (am_c)^2$, or at the level 
of~$\pm5$\,MeV. The coefficient~$c_\mathrm{hfs}$ is readily computed in 
perturbation theory and this calculation is underway.

The second correction comes from residual taste-exchange interactions, 
which from the data in Fig.~\ref{hfs-fig} could be of order a few MeV. 
Both this error and that corrected by $\delta\lag_\mathrm{hfs}$ are 
approximately proportional to~$a^2$. So we can estimate them (together) 
by comparing to a calculation with a different lattice spacing. We 
repeated our hyperfine splitting analysis on the $1/8$\,fm lattice 
using the tuned~$\epsilon$~($=-.21$).  The $a^2$~errors should be a 
little more than twice as large on the coarser lattice since $a^2$ is 
almost exactly twice as large. We obtained a splitting of 110(3)\,MeV 
on the coarse lattice, which is essentially identical to 
the~109(3)\,MeV we obtained on the fine lattice. Combining these 
results, together with our \emph{a priori} expectation of~$\pm5$\,MeV 
errors from $a^2$~corrections, we obtain an $a^2$~corrected hyperfine 
splitting is~109(5)\,MeV.

The third correction is due to the fact that our simulation does not 
include effects from the annihilation of the valence $c\overline{c}$ 
quarks into two or more gluons. Such annihilations are responsible for 
small shifts in the $\eta_c$ and $\psi$ masses, as well as for the 
(non-electromagnetic) hadronic decay rate of each meson. The dominant 
contribution comes from $c\overline{c}\to gg$ and affects only 
the~$\eta_c$. The shift in the $\eta_c$ energy is proportional to the 
perturbative amplitude for $c\overline{c}\to gg \to c\overline{c}$ at 
threshold~\cite{bodwin-braaten-gpl} and 
therefore~\cite{positronium-result}
\be
\begin{aligned}
\Delta E_{\eta_c} - & i\,\Gamma(\eta_c\to\text{hadrons})/2\\
& \propto \ln(2)-1 - i\,\frac{\pi}{2} + \order(\alpha_s).
\end{aligned}
\ee
This result implies that the leading correction to the $\eta_c$ mass 
due to $c\overline{c}$ annihilation can be computed from the hadronic 
width:
\be
\begin{aligned}
\Delta E_{\eta_c} &= \Gamma(\eta_c\to\text{hadrons})
\left(\frac{\ln(2)-1}{\pi} +\order(\alpha_s)\right) \\
	&= -2.4(8)\,\text{MeV}
\end{aligned}
\ee
where we use experimental results from~\cite{etac-data}.
This correction increases our theoretical value for the 
$\psi$--$\eta_c$ splitting to 111(5)\,MeV, which agrees well with 
experiment. The sea-quark masses in our simulations are not quite 
correct, but both theoretical expectations and experience with previous 
simulations indicate that this has negligible effect on this hyperfine 
splitting.

This level of precision would be impossible using our ASQTAD results 
because the taste-exchange errors are tens of~MeV\,---\,much larger 
than the $\pm5$\,MeV  expected from other $a^2$~errors. This example 
confirms that taste-exchange errors are likely the dominant source of 
$a^2$~errors in the ASQTAD formalism. With HISQ, on the other hand, 
taste-exchange errors have been suppressed to a level commensurate with 
other errors.

\begin{figure}
		\begin{center}
		\includegraphics[scale=1]{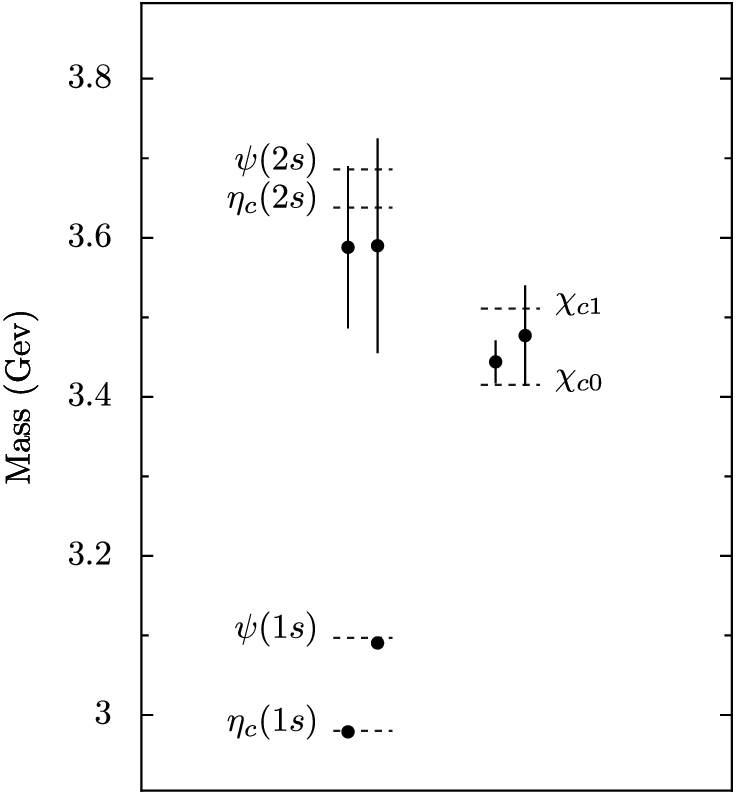}
	\end{center}
	\caption{Masses for different excitations of the $\psi$~meson from a 
simulation at lattice spacing~$1/11$\,fm. The dashed lines indicate the 
results from experiment.}
	\label{psi-spect-fig}
\end{figure}

Finally we also computed the masses of some radially and orbitally 
excited states in the $\psi$~family using HISQ, although not as 
accurately as the ground-state masses. High-precision determinations of 
excited-state masses require careful design of the meson sources and 
sinks used in the simulation. Here we did not attempt high precision, 
but rather used simple local sources to do a quick check on the 
spectrum. Our results, shown in Fig.~\ref{psi-spect-fig}, agree well 
with experiment to within our statistical errors.

\section{Simulation Techniques for Highly-Smeared Operators}
It would be highly desirable to create new unquenched gluon 
configurations using the HISQ action in place of the ASQTAD action. The 
use of such configurations, with lattice spacings of order $a=0.1$\,fm, 
would significantly reduce any residual worries about errors due to 
taste-exchange interactions.

Such simulations are complicated, however, by the heavily smeared, 
reunitarized links in the action. The additional smearing and 
reunitarization have no effect on the cost of the quark-matrix 
inversions required when updating gluon configurations, but they 
complicate calculation of the derivative~$D_a(x,\mu)\,S_q$ of the quark 
action~$S_q[U]$ with respect to an individual link operator~$U_\mu(x)$. 
Derivative~$D_a(x,\mu)$ is defined by
\be
f(\mathrm{e}^{i\epsilon^a T^a} U_\mu(x)) \equiv f(U_\mu(x)) + 
\epsilon^a D_a(x,\mu) f(U_\mu) + \order(\epsilon^2)
\ee
for any function~$f$ of~$U_\mu(x)$. We developed and tested both an 
analytic and a stochastic version of the derivative for this 
action~\cite{Follana:2003fe}.

The analytic version employs a unitary projection
\begin{equation}
V \to (V V^\dagger)^{-\frac{1}{2}} V
\end{equation}
to reunitarize each smeared link $V$.  The main obstacle is then
computation of the gauge derivative of the inverse square root.  Using 
the product-rule identity for the matrix $M\equiv V V^\dagger$,
\begin{equation}
M^{\frac{1}{2}} \left(D M^{-\frac{1}{2}}\right) +
\left(D M^{-\frac{1}{2}}\right) M^{\frac{1}{2}} =
-M^{-\frac{1}{2}} \left(D M\right) M^{-\frac{1}{2}} ,
\end{equation}
we can solve for $D (V V^\dagger)^{-\frac{1}{2}}$ directly,
both iteratively by the conjugate gradient algorithm, and exactly
by first diagonalizing $V V^\dagger$.

Using the chain rule, we combine $D (V V^\dagger)^{-\frac{1}{2}}$ with
standard derivatives of the base action and of the smeared links $V$.
We encode derivatives of the action and smeared links generically,
allowing for run-time changes independently in either.  The additional
cost of computing and combining $D (V V^\dagger)^{-\frac{1}{2}}$ with 
these
is minimal.

Several other analytic approaches have been developed for
unitarized smearings~\cite{Kamleh:2004xk, Morningstar:2003gk}.
The authors of \cite{Kamleh:2004xk} addressed the problem of computing
the derivative of the matrix inverse square root by replacing
it with a rational approximation.
In \cite{Morningstar:2003gk}, the smearing itself is unitary, 
explicitly
avoiding the need to reunitarize and the inverse square root.  In both, 
the
derivatives are then computed straightforwardly.

In the stochastic approach, we define links
\be
U^{\eta}_\mu(x) \equiv \mathrm{e}^{i\epsilon\eta(x,\mu)} U_\mu(x)
\ee
where $\eta(x,\mu)$ is now a field of traceless, hermitian $3\times3$ 
random matrices, each with normalization
\be
\langle \eta_{ij} \eta_{lm} \rangle_\eta = \left(\delta_{im}\delta_{jl} 
-
\frac{1}{3}\delta_{ij}\delta_{lm}\right),
\ee
and $\epsilon$ is a very small number. Defining
$\Delta S_q(x,\mu)$  to be the terms in $S_q[U^\eta]-S_q[U]$ containing 
$U_\mu(x)$,
the derivative can be computed efficiently from
\be
2\epsilon T^aD_a(x,\mu)\,S_q \approx \langle\eta(x,\mu) \, \Delta 
S_q(x,\mu) \rangle_\eta
\ee
averaged over a finite number of sets of random~$\eta$s. Our numerical 
experiments suggest that 10--100 sets of~$\eta$s are adequate for 
actions like~HISQ. We will describe both our stochastic and analytic 
techniques in a later paper.

\section{Conclusions}
In this paper, we have demonstrated that taste-exchange interactions 
are perturbative and we have shown how to use Symanzik improvement to 
create a new staggered-quark action (HISQ) that has greatly reduced 
one-loop taste-exchange errors, no tree-level order~$a^2$ errors, and 
no tree-level order~$(am)^4$ errors to leading order in the quark's 
velocity~$v/c$. The HISQ action addresses one of the fundamental issues 
surrounding staggered-quark simulations by allowing us to estimate 
taste-exchange interactions through comparisons of HISQ with ASQTAD 
results. We presented numerical evidence that taste-exchange 
interactions in HISQ contribute less than~1\%  to light-quark 
quantities, like meson masses and decay constants, at lattice spacings 
as large as~0.1\,fm.

The suppression of all order~$(am)^4$ errors by powers of $(v/c)^2$ 
makes HISQ the most accurate discretization of the quark action for 
simulating $c$~quarks on current lattices. We demonstrated this with a 
new lattice QCD determination of the $\psi-\eta_c$ mass splitting, 
which agrees well with experiment. This result could be improved by 
computing the coefficient $c_\mathrm{hfs}$ in the 
correction~$\delta\lag_\mathrm{hfs}$ (\eq{L-hfs}) to the HISQ action.

Our final HISQ action is defined by Eqs.~(\ref{hisq-def}--\ref{X-def}). 
We showed that the tree-level value (\eq{tuneepsilon}) for the 
Naik-term parameter~$\epsilon$ is adequate at the level of 1\%~errors 
for lattice spacings at least as large as~$1/8$\,fm.

\begin{figure}
	\begin{center}
		\includegraphics[scale=1]{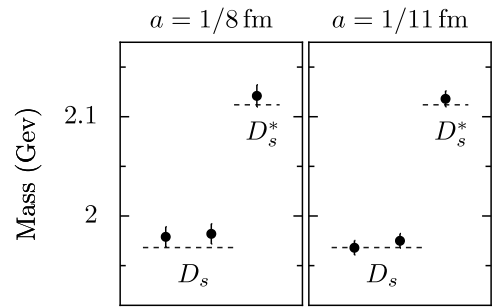}
	\end{center}
	\caption{Masses for two different tastes of the~$D_s$ and one for the 
$D_s^*$ using HISQ $c$~quarks on lattices with lattice spacings 
$a=1/8$\,fm ($\epsilon=-0.21$) and $1/11$\,fm ($\epsilon=-0.115$). For 
HISQ, $D_s$s are given (from left to right) for the 0-link and 
1-temporal-link mesons; only the 0-link splitting is given for 
the~$D_s^*$. The dashed lines indicate the results from experiment. The 
error bars shown are from the simulated values for 
$m_{D_s}-m_{\eta_c}/2$ and $m_{D_s^*}-m_{\eta_c}/2$, since these 
quantities are insensitive to the $c$~mass and the $c$~mass was tuned 
to make the simulated value for~$m_{\eta_c}$ exact.}
	\label{Dhfs-fig}
\end{figure}

The HISQ formalism will be most useful for $D$~physics and works very 
well for such mesons. In Fig.~\ref{Dhfs-fig}, for example, we show the 
$D_s^*$--$D_s$ spin splitting from simulations with HISQ $c$~quarks 
using both our $1/8$\,fm and $1/11$\,fm lattices. As in the charmonium 
case, theory and experiment agree well here to within errors. Another 
very sensitive test of our simulations is to compare computed and 
experimental values for the mass difference $2m_{D_s}-m_{\eta_c}$, 
which is quite insensitive to tuning errors in the~$c$~mass. We 
obtain~956(14)\,MeV using HISQ on the $1/11$\,fm lattice, 
and~978(20)\,MeV on the $1/8$\,fm lattice. Both values agree well with 
the splitting~956(1)\,MeV from experiment. Finally, we measured the 
speed of light and found $c^2=1.00(4)$ for the~$D_s$ on the coarse 
lattice ($a=1/8$\,fm with $\epsilon=-0.21$), confirming that the same 
coupling constant values work for both~$D_s$ and $\eta_c$~mesons. It is 
important to appreciate that there are no free parameters available for 
tuning in the extraction of any of these results; all QCD parameters 
were tuned using other quantities.

As we will discuss in a later paper, the HISQ formalism is particularly 
useful for accurate calculations of quantities like $f_D$, $f_{D_s}$, 
$D\to\pi l \nu$, and so on. These all require currents and these 
currents, even though they are conserved or partially conserved, have 
order~$\alpha_s (am)^2$ renormalizations. Consider, for example, 
$c\overline{c}$~annihilation into a (virtual) photon. The 
electromagnetic current can be computed by inserting photon link 
operators,
\be
U^\mathrm{QED}_\mu \equiv \exp\left( -ie_q\int_x^{x+\hat\mu}dx\cdot 
A^\mathrm{QED}
\right),
\ee
into the HISQ action and then expanding to first order in 
$A^\mathrm{QED}$ to obtain
\be
-i e_q \, A_\mu^\mathrm{QED} \, J_\mu^\mathrm{QED}.
\ee
Current $J^\mathrm{QED}$ is not renormalized, because of QED's gauge 
symmetry, but it is not the only operator that contributes to 
$c\overline{c}$~annihilation. In addition there are $\order(a^2)$ 
correction terms that contribute,
\be
\begin{split}
c_1\,\alpha_s\, a^2 m\, \psib & \,\sigma\cdot eF^\mathrm{QED}\,\psi \\
& + c_2\,\alpha_s\, a^2\, \psib \,(D\cdot 
eF)^\mathrm{QED}\cdot\gamma\,\psi,
\end{split}
\ee
just as for the gluon fields. The first correction term renormalizes 
the quark's magnetic moment, the second its charge radius. Such terms 
are normally negligible since the extra derivatives introduce extra 
powers of $(ap)^2$ and $p$ is small. For $c$~quark annihilations, 
however, the extra derivatives become $c$~quark masses instead of small 
momenta, and so are more important. It is easy to compute 
coefficients~$c_1$ and~$c_2$ in perturbation theory, but contributions  
to annihilation from these operators are indistinguishable from those 
coming from the leading operator~$J^\mathrm{QED}$. Consequently we can 
omit the corrections and, instead, introduce a renormalization constant 
for $J^\mathrm{QED}$:
\be
J_\mu^\mathrm{QED,continuum} = 
Z_\mathrm{eff}\,J^{\mathrm{QED},\mathrm{HISQ}}_\mu
\ee
where
\be
	Z_\mathrm{eff} = 1 + c\,\alpha_s\,(am)^2 + \order(\alpha_s^2\,(am)^2).
\ee

The situation is very similar for (partially-conserved) 
weak-interaction currents. Again the current derived from the action 
requires $\order(\alpha_s(am)^2)$~corrections. Once these one-loop 
corrections have been calculated and included, however, the remaining 
terms are only $\order(\alpha_s^2(am)^2)$. Consequently one-loop 
radiative corrections are all that is necessary to achieve 
1--2\%~precision for $f_D$, $D\to\pi l\nu$, and similar quantities. 
Other discretization errors in these quantities should be of order~1\% 
or less. With one-loop renormalizations, lattice results will be more 
accurate than current results from CLEO-c and the $b$~factories.

\section*{Acknowledgments}
We thank the MILC collaboration for sharing their configurations with 
us. This work was funded by grants from the NSF, DOE and PPARC. The 
calculations were carried out on computer clusters at Scotgrid and 
QCDOCX; we thank David Martin and EPCC for assistance.

\appendix

\section{Gamma Matrices}\label{app-gamma}
The complete set of spinor matrices can be labeled by a four-component 
vector~$n_\mu$ consisting of 0s and~1s (i.e., $n_\mu\in\mathbb{Z}_2$):
\be \label{spinors}
\gamma_n \,\equiv\, \prod_{\mu=0}^3(\gamma_\mu)^{n_\mu}.
\ee
We will also sometimes use the more conventional, but equivalent (up to 
a phase) hermitian set:
\begin{gather}
1,
\quad\gamma_5\equiv\gamma_t\gamma_x\gamma_y\gamma_z,
\quad\gamma_\mu, \nonumber \\
\quad \gamma_{5\mu}\equiv i\gamma_5\gamma_\mu,
\quad\gamma_{\mu\nu}\equiv\ihalf [\gamma_\mu,\gamma_\nu].
\end{gather}
The $\gamma_n$s have several useful properties:
\begin{itemize}
\item orthonormal:
	\be
	\Tr(\gamma_n^\dagger \gamma_m) = 4\,\delta_{nm}
	\ee
\item closed under multiplication:
	\be
	\gamma_n\gamma_m = (-1)^{n\cdot m^<}\,\gamma_{n+m}
	\ee
where
	\be
	m_{\mu}^< \equiv \sum_{\nu<\mu} m_\nu \bmod 2
	\ee
and
	\be
	n\cdot m^< = n^>\cdot m
	\ee
with
	\be
	n^>_{\mu} \equiv \sum_{\nu>\mu} n_\nu \bmod 2;
	\ee
\item hermitian or antihermitian:
	\be \label{gammadag}
	\gamma_n^\dagger = (-1)^{n\cdot n^<}\,\gamma_n = \gamma_n^{-1};
	\ee
\item commuting or anticommuting:
	\be
	\label{commrel}
	\gamma_n \,\gamma_m = (-1)^{\overline{m}\cdot n}\,\gamma_m\,\gamma_n
	\ee
where
	\bea
	\overline{m}_\mu&\equiv&  m^>_\mu + m^<_\mu  =
	\sum_{\nu\ne\mu} m_\nu \bmod 2  \nl
	&=&
	\begin{cases}
		m_\mu & \text{if $m^2$  even} \cr
		(m_\mu+1) \bmod 2 & \text{if $m^2$ odd,}\cr
	\end{cases}\label{mbar}
	\eea
and
	\be
	\overline{m}\cdot n = \overline{n}\cdot m,
	\ee
	\be
	\overline{\overline{m}}=m \qquad (\overline{m}\cdot m)\bmod2 = 0;
	\ee
\item permutation operator: if one uses the standard representation for 
gamma matrices, where
\be
\gamma_0 = \left(\begin{array}{cc} 1 & 0 \\ 0 & -1 
\end{array}\right)\quad\quad
\gamma_i = \left(\begin{array}{cc} 0 & \sigma_i \\ \sigma_i & 0 
\end{array}\right)
\ee
and
\begin{gather}
\sigma_1 = \left(\begin{array}{cc} 0 & 1 \\ 1 & 0 
\end{array}\right)\quad\quad
\sigma_2 = \left(\begin{array}{cc} 0 & -i \\ i & 0 
\end{array}\right)\quad\quad
\nonumber \\
\sigma_3 = \left(\begin{array}{cc} 1 & 0 \\ 0 & -1 \end{array}\right),
\end{gather}
then there is at most one nonzero element in any row or any column of 
any $\gamma_n$, and that element is $\pm 1$ or $\pm i$. Thus 
multiplying a spinor~$\psi$ by $\gamma_n$ simply permutes the spinor 
components of $\psi$, multiplying each by $\pm 1$ or $\pm i$.
\end{itemize}
A convenient notation, reminiscent of $\gamma_{5\mu}$, is
\be
\gamma_{nm} \equiv \gamma_n\,\gamma_m .
\ee

\section{Staggered Quarks}\label{about-staggered-quarks}
The naive discretization of the quark action is formally equivalent to 
the staggered-quark discretization. Staggering is an important 
optimization in simulations; it is also a remarkable property. Consider 
the following local transformation of the naive-quark field:
\be \label{chidef}
\psi(x) \to \Omega(x)\,\chi(x) \quad\quad \psib(x)\to\chib(x)\,
\Omega^\dagger(x)
\ee
where
\be
\Omega(x) \equiv \gamma_{x} \equiv \prod_{\mu=0}^{3} 
(\gamma_\mu)^{x_\mu},
\ee
and we have set the lattice spacing~$a=1$ for convenience. (We will use 
lattice units, where $a=1$, in this and all succeeding appendices.)
Note that
\be
\Omega(x) = \gamma_n \quad\quad\text{for ${n_\mu=x_\mu\,\bmod\,2}$;}
\ee
there are only 16~different~$\Omega$s. It is easy to show that
\bea
\alpha_\mu(x)&\equiv&\Omega^\dagger(x)\gamma_\mu\Omega(x\pm\hat\mu) 
\,=\,
(-1)^{x^<_{\mu}}, \\  %
1 &=& \Omega^\dagger(x)\Omega(x)
\eea
where $x^<_{\mu} \equiv x_0+x_1+\cdots+x_{\mu-1}$ (see 
Appendix~\ref{app-gamma}).
Therefore the naive-quark action can be rewritten
\be\label{staggaction}
\psib(x)(\gamma\cdot\Delta+m)\psi(x) \,=\,
\chib(x)(\alpha(x)\cdot\Delta + m)\chi(x).
\ee
Remarkably the $\chi$~action is diagonal in spinor
space; each component of $\chi$ is exactly equivalent to every other
component. Consequently the $\chi$ propagator is diagonal in spinor
space in \emph{any background gauge field}:
\be
\langle\chi(x)\chib(y)\rangle_\chi \,=\,s(x,y)\,{\bf 1}_{\rm spinor}
\ee
where $s(x,y)$ is the one-spinor-component staggered-quark
propagator.  Transforming
back to the original naive-quark field we find that
\be \label{propagator}
S_F \,\equiv\,\langle \psi(x)\psib(y)\rangle_\psi
\,=\,s(x,y)\,\Omega(x)\Omega^\dagger(y).
\ee

This last result is a somewhat surprising consequence of the doubling 
symmetry.
It says that the spinor structure of
the naive-quark propagator is completely independent of the gauge
field. This is certainly not the case for individual tastes of naive
quark, whose spins will flip back and forth as they scatter off
fluctuations in the chromomagnetic field, for example.
The sixteen tastes of the naive
quark field are packaged in such a way, however, that all gauge-field
dependence vanishes in the spinor structure.

Doubling symmetry is immediately evident in the staggered action, 
\eq{staggaction},
since the action is invariant under
\begin{equation}
	\chi(x) \to \Bz(0)\,\chi(x),
\end{equation}
which merely scrambles the (equivalent) spinor components of $\chi(x)$, 
and
since
\begin{equation} \label{B-O-comm}
	\Bz(x)\,\Omega(x) = \Omega(x)\,\Bz(0).
\end{equation}
In simulations one generally discards all but one spinor component of 
$\chi(x)$, resulting in highly efficient algorithms. The Symanzik 
improvements discussed above are trivially incorporated.

Note, finally, that \eq{B-O-comm} implies
\be
\Bz(x)\,\Omega(x) \Omega^\dagger(y)\,\Bz^\dagger(y)
\,=\,\Omega(x) \Omega^\dagger(y).
\ee
and therefore the naive-quark propagator, $S_F$, satisfies
\be
S_F(x,y) \,=\,  \Bz(x)\,S_F(x,y)\,\Bz^\dagger(y)
\ee
for {any background gauge field.} In momentum space this becomes
\be \label{SFdoubling}
S_F(p,q) \,=\,
\Bz(0)\,S_F(p\!+\!\zeta\pi,q\!+\!\zeta\pi)\,\Bz^\dagger(0),
\ee
which is an exact relationship that is useful in perturbative
calculations. This last relation, which is easily checked at tree level 
(but
true to all orders in~$\alpha_s$), shows that there is only one
sixteenth as much information in the naive-quark propagator as naively
expected.

\section{Taste, Naive vs.\ Staggered}\label{naive-staggered-psi}
The quark field $\psi(x)$ will typically contain contributions from all 
16~tastes. One can separate out (approximately) the different tastes  
by blocking the field on hypercubes that have two sites per side. One 
way to project out the $\zeta=0$ taste, for example, is to average over 
the hypercube:
\be
\psi_B = \frac{1}{16}\sum_{\delta x_\mu\in\mathbb{Z}_2} \psi(x_B+\delta 
x)
\ee
where $x_B$, with $x_{B\mu} \bmod 2 = 0$, identifies the hypercube, and 
the sum is over all 16~sites in the hypercube. Any component of 
$\psi(x)$ that has momentum $p\approx\zeta\pi$ with $\zeta\ne0$ will be 
strongly suppressed by the average. (The suppression is 
$\order(a^2p^2)$ for a mode with momentum $p+\zeta\pi$.) A $\zeta\ne0$ 
piece of $\psi(x)$ can be isolated by applying a doubling 
operator~$\Bz$ to transform that component to $\zeta=0$, and then, 
again, averaging over the hypercube to isolate that component:
\be
\label{psiB}
\psi_B^{(\zeta)} = \frac{1}{16}\sum_{\delta x_\mu\in\mathbb{Z}_2} 
\Bz(x_B+\delta x) \, \psi(x_B+\delta x).
\ee
The 16~blocked fields $\psi_B^{(\zeta)}$, with one for each~$\zeta$, 
describe the 16~different tastes of quark, each now in the low momentum 
sector (i.e., with $p_\mu\le\pi/2$).

With staggered quarks, one keeps only one component of $\chi(x)$ since 
the components are equivalent and decouple:
\be
\chi(x) \to \left[ \begin{array}{c} \chi_1(x) \\ 0 \\ 0 \\ 0 
\end{array} \right].
\ee
We can use our hypercube blocking (\eq{psiB}) to translate this single 
field back into blocked fields for ``ordinary'' quarks of different 
tastes~$\zeta$:
\be \label{stagg-naive}
\begin{split}
\sum_{\deltax_\mu\in\mathbb{Z}_2}
					&	\Bz(x_B+\deltax) \,\psi(x_B+\deltax) \\
= &
\sum_\deltax \Omega(\deltax) \, \Bz(0) \left[ \begin{array}{c} 
\chi_1(x_B+\deltax) \\ 0 \\ 0 \\  0 \end{array} \right].
\end{split}
\ee
Only four of the sixteen blocked fields are independent, however, 
because $\Bz(0)\,\chi(x)$ is always proportional to one of only four 
different spinors:
\be
\left[ \begin{array}{c} \chi_1(x) \\ 0 \\ 0 \\  0 \end{array} \right], 
\:
\left[ \begin{array}{c} 0 \\ \chi_1(x) \\ 0 \\  0 \end{array} \right], 
\:
\left[ \begin{array}{c} 0 \\ 0 \\ \chi_1(x) \\ 0 \end{array} \right], 
\:
\left[ \begin{array}{c} 0 \\ 0 \\  0 \\ \chi_1(x) \end{array} \right].
\ee
(This is because $\Bz(0)=\gamma_{\overline{\zeta}}$ and all $\gamma_n$s 
when applied to a spinor merely permute the elements of the spinor, 
multiplying each by $\pm 1$ or $\pm i$.) Consequently we can 
reconstruct just four tastes, $t=1\ldots4$, of blocked quark from a 
single staggered field:
\be \label{stagg-taste}
\tilde{\psi}_B^{(t)} = \sum_\deltax 
\Omega(\deltax)\,\hat{\chi}^{(t)}\,\chi_1(x_B+\deltax)
\ee
where the $\hat{\chi}^{(t)}$ are unit spinors with $\hat{\chi}^{(t)}_i 
= \delta_{i,t}$.

This last formula defines the standard taste basis for staggered 
quarks. It is of more use conceptually than practically, and we will 
not need it in what follows.

Formula~(\ref{stagg-naive}) shows that some $\zeta$'s become 
indistinguishable in the staggered approximation. In fact, by explicit 
calculation (given our specific representation of the 
$\gamma$~matrices), one finds that the $\zeta$s fall into four 
equivalence classes composed of indistinguishable~$\zeta$s:
\bea
A: && 0000 \quad 0001 \quad 0110 \quad 0111 \nonumber\\
B: && 0010 \quad 0011 \quad 0100 \quad 0101 \nonumber\\
C: && 1000 \quad 1001 \quad 1110 \quad 1111 \nonumber\\
D: && 1010 \quad 1011 \quad 1100 \quad 1101
\eea
These four classes correspond to the four tastes of staggered quark; 
all the $\zeta$s in a single class give the same staggered-quark 
field~$\tilde{\psi}_B^{(t)}$ (\eq{stagg-taste}) from~\eq{stagg-naive}.

The equivalence of the $\zeta$s within a single class is preserved 
under addition of $\zeta$s: for example, adding any two $\zeta$s from 
class~$B$ always gives a $\zeta$ in class~$A$, while adding any vector 
from class~$B$ to any vector from class~$C$ always gives a vector in 
class~$D$. The full addition table for these classes is:
\begin{center}
\begin{tabular}{c|cccc}
 $+$   & $A$ & $B$ & $C$ & $D$ \\ \hline
$A$ & $A$ & $B$ & $C$ & $D$ \\
$B$ & $B$ & $A$ & $D$ & $C$ \\
$C$ & $C$ & $D$ & $A$ & $B$ \\
$D$ & $D$ & $C$ & $B$ & $A$
\end{tabular}
\end{center}
This kind of structure is necessary for the four-fold reduction in the 
number of tastes due to staggering\,---\,that is, 16 tastes of naive 
quark are reduced to 4~tastes of staggered quark because we can 
consistently identify certain corners of the Brillouin zone with each 
other in the staggered case.

\section{Naive-Quark Currents}
Naive quarks lead to huge numbers of nearly equivalent mesons. Where in 
ordinary QCD one might consider a single meson operator, 
$J_n=\psib\gamma_n\psi$, one has 16~point-split operators in the naive 
theory each of which couples to a different meson:
\bea
J_n^{(s)}(x) &\equiv& \psib(x)\,\gamma_n^{(s)}\,\psi(x) \nl
&\equiv& \psib(x)\,\gamma_n\,\psi(x+\delta x_{sn}) \nl
&\propto& \chib(x)\,\gamma_s\,\chi(x+\delta x_{sn}),
\eea
where $s$ is any one of the 16~four-vectors consisting of~0s and~1s 
only,
\be
\delta x_{sn} \equiv (s+n)\,\bmod\,2  ,
\ee
and link operators~$U_\mu$ are implicit (and $a=1$~still).
Each of these operators creates a different version of the $J_n$~meson; 
they are orthogonal. This is because
\be
J_n^{(s)} \,\to\, (-1)^{s_\mu}\,J_n^{(s)}
\ee
under a doubling transformation where
\be
\psi(x) \,\to\,\B_{\zeta=\hat\mu}(x)\,\psi(x)
\qquad
\psib(x) \,\to\,\psib(x)\B_{\hat\mu}^\dagger(x).
\ee
Thus the four-vector~$s$ determines the bilinear's transformation 
properties under arbitrary doubling transformations; it specifies the 
bilinear's ``signature'' under doubling transformations.
Since doubling transformations are symmetries of the naive theory, the 
doubling signature is conserved: for example,
\be
\langle\,J_m^{(r)}\,\,J_n^{(s)\dagger}\,\rangle
\,=\,0\quad\quad\mbox{if $r\ne s$},
\ee
which proves that each of our point splittings creates a different 
meson.

Different signatures correspond to different variations of the same 
continuum meson, typically with slightly different masses, etc. These 
are the different tastes of the meson. We label different tastes by the 
corresponding signature~$s$ in the meson's rest frame.

Additional mesons are made by boosting particles into other corners of 
the Brillouin zone:
\be
 p_\tot \approx \zeta\pi
\ee
for one of the sixteen~$\zeta$s. Such mesons would be highly 
relativistic in the continuum, but here they are equivalent to 
low-energy mesons because of the doubling symmetry: for example, the 
antiquark in the meson might carry a momentum near zero, while the 
quark carries $\zeta\pi$ but is then equivalent to a zero-momentum 
quark state through the doubling symmetry. Pushing such momenta through 
a naive-quark bilinear,  using for example
\be
\sum_x \mathrm{e}^{i\,p_\tot\cdot x}\,J^{(s)}_n(x),
\ee
changes the quantum numbers of the meson created by the operator. To 
see this, consider current
\bea \label{j-n-zeta-s}
J_n^{(\zeta,s)} (x) &\equiv&
(-1)^{\zeta\cdot x}\,\psib(x)\,\gamma_{\,\overline{\zeta}}^\dagger\,
\gamma_{n}^{(s)}\,
\psi(x) \nl
&=&
\psib(x)\,\Bz^\dagger(x)\,\gamma_n^{(s)}\,\psi(x)\rule{0pt}{3ex}\\
&\propto& 
\chib(x)\,\gamma_{\,\overline\zeta}^\dagger\,\gamma_s\,\chi(x+\delta 
x_{sn})
\rule{0pt}{3ex}
\eea
which has signature $(\overline{\zeta}+s)\bmod 2$.
This current obviously carries momentum $p_\tot = \zeta\pi$ if averaged 
over all~$x$. It is easy to show that flavor-nonsinglet mesons created 
by~$J_n^{(\zeta,s)}$ with different~$\zeta$s are all identical, and 
therefore all carry taste~$s$. For a $\overline{u}d$~meson, for 
example, we can use the \emph{separate} doubling symmetries of the $u$ 
and $d$ to prove that
\bea
\langle\,
J_n^{(\zeta,s)}(x)\,
J_n^{(\zeta,s)\dagger}(y)
\,\rangle
&\equiv& \langle\,\overline u\Bz^\dagger \gamma_n^{(s)}d\,\,
\overline d \gamma_n^{(s)}\Bz u\,\rangle \nl
&=& \langle\,\overline u\gamma_n^{(s)} d\,\,
\overline d \gamma_n^{(s)}u\,\rangle \rule{0pt}{3ex}\nl
&\equiv&
\langle\,
J_n^{(s)}(x)\,
J_n^{(s)\dagger}(y)
\,\rangle \rule{0pt}{3ex}
\eea
for all~$\zeta$s.

The labeling on~$J_n^{(\zeta,s)}$ is intuitive (and therefore useful) 
only if the current is averaged over~$x$ in such a way that~$p_\tot = 
\zeta\pi + p$ where $-\pi/2<p_\mu\le\pi/2$ for all~$\mu$; the 
combination of all $\zeta$s and $p$s covers all of momentum space, so 
nothing is lost by this restriction and double counting is avoided. We 
can enforce this restriction on the momenta by redefining the current 
on a blocked lattice with one site at the center of every 
$2^4$~hypercube on the original lattice, just as we did for the quark 
field (Section~\ref{naive-staggered-psi}):
\be
J_{Bn}^{(\zeta,s)} \equiv \sum_{\delta x_\mu\in\mathbb{Z}_2}
J_n^{(\zeta,s)}(x_B +\delta x)
\ee
where $x_B$ identifies the hypercube (with $x_{B\mu}\bmod2 = 0$). The 
blocked current creates a meson of taste~$s$ with momentum in the 
$\zeta\pi$ corner of the Brillouin zone.

Note that momentum and taste conservation imply
\be
\sum_{x,y}
\langle\,
J_n^{(\zeta,s)}(x)\,
J_n^{(\zeta^\prime,s^\prime)\dagger}(y)
\,\rangle
\,=\,0
\ee
unless $\zeta=\zeta^\prime$ and $s=s^\prime$. Consequently the 
different operators are orthogonal when analyzed in momentum space. 
Typically we sum over space, however, but not time. In that case 
operators with the same signature, but where 
$\zeta^\prime-\zeta=\pm\hat t$, can mix. This mixing leads to 
components in meson propagators that oscillate in time, as we discuss 
below.

To summarize, there are 16~sets, each set labeled by taste~$s$, of 
16~identical mesons, each meson labeled by~$\zeta$, for each 
flavor-nonsinglet meson in the continuum. Each corner~$\zeta$ of the 
Brillouin zone has a single representative of each taste of meson.

Flavor-singlet mesons are slightly more complicated. In a 
$\overline{u}u$~meson, for example, the quark and antiquark are created 
by the same field, and so do not have separate doubling 
transformations. Therefore the argument relating $J_n^{(\zeta,s)}$s 
with different $\zeta$s doesn't work. The only contributions that spoil 
this argument are from annihilation, where the meson's quark and 
antiquark annihilate into gluons. Annihilation gluons that contribute 
to, for example,
\be
\sum_{\vec x}\langle\,
J_n^{(\zeta,s)}(x)\,
J_n^{(\zeta,s)\dagger}(0)
\,\rangle
\ee
carry total momentum~$\zeta\pi$ and so are far off-shell unless 
$\zeta=0$. This has two implications: First, annihilation contributions 
will be different for different $\zeta$s. And, second, only the 
$\zeta=0$ case has the correct coupling between the flavor-singlet 
quarks and purely gluonic channels. In fact, only the taste-singlet 
state, among the $\zeta=0$ states, couples to the gluons since
\bea
\langle\,J_n^{(s)}(x)\,\rangle_\psi &\propto& \Tr(\gamma_s) \nl
&=& 0 \quad\quad \mbox{unless $s=0$}.
\eea

A similar condition applies to nonzero~$\zeta$s as well. For 
each~$\zeta$ there is only one taste that can couple to gluons. The 
gluons are highly virtual if $\zeta$ is nonzero. These last 
contributions are taste-violating because they change quark taste along 
quark lines; they are removed by the contact terms discussed in this 
paper.

It is not surprising that the flavor-singlet mesons are more 
complicated. They usually have to be, particularly in the pseudoscalar 
channel where the $U(1)$~problem must be resolved. In our naive-quark 
theory, only the $\zeta=s=0$ mesons couple properly to gluons. The 
masses of pseudoscalars with $\zeta=s=0$ are  shifted properly by 
instantons in the chiral limit, so that the $U(1)$~problem is resolved. 
The $\zeta=s=0$~neutral pion is also the only pion that decays to two 
photons. The corresponding axial-vector current is only approximately 
conserved, even in the chiral limit, so anomalies are not needed to 
mediate the photon decay.

\section{Staggered-Quark Currents}\label{app-stagg-currents}
The 256 different mesons created by the naive-quark 
bilinears~$J_n^{(\zeta,s)}$ include 16~identical copies of each 
distinct taste of meson. Staggering the quark fields discards identical 
copies, leaving just 16~distinct tastes. The naive-quark bilinears 
corresponding to the staggered-quark bilinears are the ones that become 
diagonal after they are staggered. Any other operator creates mesons 
that mix different spinor components of the staggered quark 
operator~$\chi(x)$ (\eq{chidef}), and so is discarded when we stagger.

We can identify the bilinears that survive staggering by noting that
\bea
J^{(\zeta,s)}_n &\propto& 
\chib\,\gamma_{\,\overline\zeta}^\dagger\,\gamma_s\,\chi
\nl
&\to& \chib \,\chi
\eea
after staggering provided $\zeta = \overline{s}$.
The staggered-quark operator is therefore
\bea \label{staggop}
J_n^{(\overline{s},s)}
 &\equiv &
{(-1)^{\overline{s}\cdot x}}\,\,
\psib(x)\,\gamma_{s}^\dagger\,\gamma_{n}\,\psi(x+\delta x_{sn}) 
\nonumber \\
&=& \beta_n^{(s)}(x) \,\, \chib(x)\, \chi(x+\delta x_{sn})
\eea
where again $\delta x_{sn} = (s+n) \bmod 2$ (with $a=1$), and
\be
\begin{aligned}
\beta_n^{(s)}(x) &= \quarter\,\Tr\left(\gamma^\dagger_s\gamma^\dagger_x
	\gamma_n\gamma_{x+s+n}\right) \\
	&= (-1)^{x\cdot(s^<+n^>)}\,(-1)^{n\cdot(s+n)^<}.
\end{aligned}
\ee
Each taste~$s$ in the staggered theory corresponds to a specific corner 
of the Brillouin zone, with $p_\tot\approx\overline{s}\pi$, in the 
naive-quark theory. These operators have zero signature and so are 
unchanged under doubling transformations (\eq{dsym}).

It is useful for formal analyses, though less so for simulations, to 
introduce a slightly different definition for these bilinears by 
defining a new operator on our naive-quark field:
\be \label{tastenotn}
\gamma_n\dprod\xi_s\,\psi(x) \,\equiv\,
(-1)^{\overline{s}\cdot x} 
\,\gamma_s^\dagger\,\gamma_n\,\psi(x\oplus(n+s)),
\ee
where $\oplus$ adds vector $n+s$ to $x$ ``modulo'' the hypercube that 
$x$ lies in\,---\,that is,
\be
(x \oplus n)_\mu \equiv x_{B\mu} + (x_\mu-x_{B\mu}+n_\mu) \bmod 2
\ee
when $x$ is in the hypercube labeled by site~$x_B$ (with $x_{B\mu} 
\bmod 2 = 0$).
Thus $x\oplus n$ is in the same hypercube as~$x$. With this definition, 
we can redefine  the staggered-quark current with spin~$n$ and 
taste~$s$ to be:
\be \label{staggop-var}
\psib(x)\,\gamma_n\dprod\xi_s\,\psi(x).
\ee

The $\oplus$~means that the new operators~$\gamma_n\dprod\xi_s$ have a 
simple algebra. If, for example,
\be
\tilde{\psi}(x) \equiv \gamma_n\dprod\xi_s\,\psi(x)
\ee
then, using \eq{commrel},
\be
\gamma_m\dprod\xi_r\,\tilde{\psi}(x) = 
\gamma_m\gamma_n\dprod\xi_r\xi_s\,\psi(x),
\ee
which implies that in general
\be
\gamma_m\dprod\xi_r\,\,\gamma_n\dprod\xi_s \,\,
= \,\, \gamma_m\gamma_n \dprod \xi_r\xi_s.
\ee
Note that these definitions also imply that $\xi$s anticommute just as 
$\gamma$s: for example,
\be
\gamma_n \dprod\xi_r\xi_s = (-1)^{\overline{r}\cdot s}\,
\gamma_n \dprod\xi_s\xi_r.
\ee

We can effectively restrict our current to the $\zeta=\overline{s}$ 
corner of the Brillouin zone by again blocking on $2^4$~hypercubes to 
obtain
\be
\cJ n s (x_B) = \frac{1}{16}\sum_{\delta x_\mu\in\mathbb{Z}_2}
	\psib(x_B+\delta x)\,\gamma_n\dprod\xi_s\,\psi(x_B+\delta x)
\ee
where $x_{B\mu}\bmod 2 = 0$.
This formula is related to a more standard formula by staggering the 
meson operator:
\be
\begin{split}
\psib(x)\,\gamma_n & \dprod\xi_s\,\psi(x) \\
& = \chib(x)\,\gamma_s^\dagger\gamma_{x}^\dagger\,\gamma_{n}\,
\gamma_{x+\delta x_{sn}}\,\chi(x\oplus\delta x_{sn})
\end{split}
\ee
where the shift~$\delta x_{sn}$ guarantees that the product of 
$\gamma$s is proportional to the unit matrix. We can rewrite this in 
terms of a sum over $\delta x$s in the positive unit hypercube:
\be
\sum_{\delta x_\mu\in\mathbb{Z}_2}
\chib(x)\,\quarter
\Tr\left(\gamma^\dagger_s\,\gamma^\dagger_x\,
\gamma_n\,\gamma_{x+\delta x}\right)\,\chi(x\oplus\delta x).
\ee
Averaging over the hypercube gives a standard formula for $\cJ n s$:
\be
\frac{1}{16}\sum_{\delta x,\delta x'} \chib(x_B+\delta x) 
\mbox{$\frac{1}{4}$} \Tr\left(\gamma^\dagger_s\,\gamma^\dagger_{\delta 
x}\,
\gamma_n\,\gamma_{\delta x'}\right)\,\chi(x_B+\delta x').
\ee

Taste and spinor structure in $\psib(x)\gamma_n\dprod\xi_s\psi(x)$ are 
both described by the same kind of four vector consisting of~0s and~1s. 
For this reason it is common practice to use the same terminology for 
describing taste as we do for spinor structure. Thus, for example, 
$\psib(x)\,\gamma_5\dprod\xi_5\,\psi(x)$ creates a pseudoscalar meson 
with ``pseudoscalar taste.'' In this case $n=s=(1,1,1,1)$, and, from 
\eq{staggop},
\bea
\psib(x)\,\gamma_5\dprod\xi_5\,\psi(x) &=& 
\beta_5^{(5)}(x)\,\chib(x)\chi(x)
\nonumber \\
&=& (-1)^{x\cdot x}\,\chib(x)\chi(x).
\eea
A different pion is created by 
$\psib(x)\gamma_5\dprod\xi_{5\mu}\psi(x)$, this one with axial-vector 
taste:
\be
\psib(x)\,\gamma_5\dprod\xi_{5\mu}\,\psi(x) =
 \beta_5^{(5\mu)}(x)\,\chib(x)\chi(x\oplus\hat\mu).
\ee
The pion created by this operator is sometimes called a $5\otimes5\mu$ 
or ``1-link pion'' since the operator is split by one link or lattice 
spacing. Similarly $\psib(x)\gamma_5\dprod\xi_5\psi(x)$ creates a 
$5\otimes5$ or 0-link pion.

Other naive-quark operators can be recast in terms of the spinor/taste 
operators. For example, the naive-quark action has a chiral symmetry in 
the massless limit under transformations
\bea
\psi(x) &\to& \exp(i\,\theta\,\gamma_5\dprod\xi_5)\,\psi(x) \nl
\psib(x) &\to& \psib(x)\,\exp(i\,\theta\,\gamma_5\dprod\xi_5),
\eea
where the~$\xi_5$s express the fact that the symmetry operation does 
not translate (move)~$\psi(x)$.

\section{One Loop Taste-Changing}\label{app-oneloop}
The Symanzik procedure for removing $\order(a^2)$ one-loop 
taste-exchange effects from the ASQTAD action involves two steps: 1) we 
compute taste-changing amplitudes for $q\overline{q}\to q\overline{q}$ 
with massless quarks in one-loop order using  lattice perturbation 
theory; and 2) we design local taste-changing counterterms for the 
staggered-quark action that cancel these one-loop amplitudes. We find 
it easiest to work with the naive-quark theory, converting to staggered 
quarks only at the end.

In order~$a^2$ we need only consider quarks at threshold\,---\,that is, 
quarks with momentum $p=\pm\zeta\pi$ for one of the sixteen $\zeta$'s 
with $\zeta_\mu\in\mathbb{Z}_2$ (where, again,~$a=1$). Taste-changing 
amplitudes are ones where the incoming and outgoing momenta along any 
particular quark line differ by $\zeta\pi$ for one of the~$\zeta$s. For 
$q\overline{q}\to q\overline{q}$, overall momentum conservation demands 
that the opposite change occur along the other quark line. Consequently 
while taste changes along individual quark lines, total taste is 
conserved. It is enough to compute amplitudes $A(0,0 ; 
\zeta\pi,-\zeta\pi)$ where the initial quarks have zero momentum and 
the final quarks have momenta $\zeta\pi$ for any~$\zeta$. (Recall that 
$\zeta\pi$ and $-\zeta\pi$ are the same momentum on the lattice.) 
Amplitudes with other quark tastes in the initial state are related to 
these amplitudes by applying the doubling symmetry to each quark line 
separately:
\begin{equation} \label{dsym-T}
	A(\zeta^\prime\pi,\zeta^{\prime\prime}\pi ;
	 		(\zeta^\prime+\zeta)\pi,(\zeta^{\prime\prime}-\zeta)\pi)
	= A(0,0;\zeta\pi,-\zeta\pi)
\end{equation}
where $\zeta^\prime_\mu,\zeta^{\prime\prime}_\mu\in\mathbb{Z}_2$.

Tree-level contributions come from Fig.~\ref{treelevel-fig}, but these 
vanish (by design) in the ASQTAD action because of ASQTAD's quark-gluon 
vertex. The only one-loop contributions that are nonzero are those 
shown in Fig.~\ref{oneloop-fig}; all other one-loop amplitudes have at 
least one quark line with only one gluon attached, and these vanish, 
again, because of ASQTAD's quark-gluon vertex. The internal quarks and 
gluons in the remaining diagrams are all highly virtual, and therefore 
these contributions can be canceled by a sum of four-quark counterterms 
each consisting of a product of two quark bilinears (with one bilinear 
per quark line).

The types of quark bilinear that can arise from these diagrams are 
greatly restricted by color conservation, chiral symmetry, and the 
doubling symmetry of the lagrangian. Quark bilinears can only carry 
singlet or octet color: thus color structure is either $\psib\psi$ or 
$\psib T^a\psi$ where $T^a$ is an $SU(3)$~generator in the fundamental 
representation. The standard chiral symmetry of the naive-quark action 
in the massless limit implies that only vector and axial-vector 
bilinears can arise in that limit: thus spinor structure is either 
$\psib\gamma_\mu\psi$ or $\psib\gamma_{5\mu}\psi$. Finally doubling 
symmetry (\eq{dsym-T}) requires that the bilinears in the 
counter\-terms must be point-split so that they are invariant under any 
doubling transformation of their quark fields (\eq{dsym}): that is, we 
have
\be
 \psib(x)\gamma_n\psi(x+\delta x(n))
\ee
or
\be
 \psib(x)\gamma_n T^a\psi(x+\delta x(n))
\ee
where~$n$ is~$\mu$ or~$5\mu$, and $\delta x(n)$ is chosen to make the 
bilinear invariant under $\psi \to \Bz \psi$ for all~$\zeta$. This last 
restriction implies that the bilinears must be staggered-quark 
operators $\cJ s n$, as defined in Appendix~\ref{app-stagg-currents}.

The one-loop amplitude, $A(0,0;\zeta\pi,-\zeta\pi)$, for a 
given~$\zeta$ is canceled by a specific set of counterterms,
\be
\frac{1}{2}\,\sum_n d^{(s)}_n\, |\cJ s n|^2 + \text{color octet 
version}
\ee
where $s=\overline\zeta$ (from \eq{staggop}). Consider, for example, 
the case where $\zeta_\mu=1$ for some~$\mu$ while~$\zeta_\nu=0$ for 
all~$\nu\ne\mu$. The taste is $s=\overline{\zeta}$ or~$5\mu$. The 
spin~$n$ depends upon the spinor structure of~$A$. For this $\zeta$ 
there are four different spinor structures, each corresponding to a 
different staggered-quark operator:
\bea
(\psib \gamma_\mu \psi)^2 &\to (\cJ {5\mu} 5 ) ^2 \nonumber \\
(\psib \gamma_\nu \psi)^2 &\to (\cJ {5\mu} {5\nu\mu} ) ^2 \nonumber \\
(\psib \gamma_{5\mu} \psi)^2 &\to (\cJ {5\mu} {1} ) ^2 \nonumber \\
(\psib \gamma_{5\nu} \psi)^2 &\to (\cJ {5\mu} {\mu\nu} ) ^2.
\eea
Each of these operators comes in color octet and color singlet 
versions, so eight counterterms are required to cancel~$A$ 
for~$\zeta^2=1$.

The coefficients~$d^{(s)}_n$ in the counterterms are computed by 
on-shell matching of the scattering 
amplitude~$A(0,0;\zeta\pi,-\zeta\pi)$ to the sum of counterterms for 
each~$\zeta$. Our results are summarized in Table~\ref{d-table}. Only 
the bilinears shown in \eq{e:tastechange} are needed here. The chiral 
perturbation theory devised by Lee and Sharpe~\cite{lee-sharpe} has ten 
additional bilinears, but eight of these are not invariant under 
doubling transformations for each quark line \emph{separately}, and two 
do not involve taste exchange.

\section{Meson Propagators}\label{app-meson-prop}
The fact that the single naive-quark field encodes sixteen different,
identical tastes of quark has practical implications for simulations. 
For example, the operator
\be
J_5(x) \equiv \psib(x)\gamma_5\Psi_b(x),
\ee
where $\Psi_b$ is an unstaggered $b$-quark
field and $\psi$ is a light-quark field,
couples only to $J^P=0^-$ mesons in the continuum, including the~$B$.
If $\psi$ is a staggered field, however, it also couples to 
$0^+$~mesons~\cite{wingate}.

The doubling-symmetry formula, \eq{dsym}, is useful in
decoding such situations. The second contribution from $J_5$ arises
from high-energy states that couple to it.  In simulations one
normally forms correlators like
\be
G_{55}(t) \,\equiv\, \sum_\xv \langle0| J_5(\xv,t) J_5^\dagger(0,0) 
|0\rangle,
\ee
where the sum over $\xv$ guarantees that total three momentum
$\pv=0$. The operators, however, are not smeared in time, and so can
create arbitrarily high-energy states. The $b$-quark
resists large energies, as these drive it far off shell, but the
staggered quark is on-shell when its energy $E\approx0$ or when
$E\approx\pi$.  These two possible states of the light quark
correspond to two different tastes. Consequently $J_5$ couples to two
different mesons: one whose light quark has taste $\zeta=0$, and one
whose light quark has taste $\zeta=(1,0,0,0)$ or $E\approx\pi$.

The first of these two meson states is the normal $0^-$~$B$~meson. To
interpret the second state, we transform the high-energy
staggered-quark field back to a low-energy field (which we understand,
since it behaves normally) using the doubling symmetry formula,
\eq{dsym}:
\be
\left.\psib(x)\right|_{E\approx\pi} \to  \psib(x) (i\gamma_5\gamma_0) 
(-1)^{t}.
\ee
Substituting in $J_5$ we see that this component of the operator is
\be
\left.\psib\gamma_5\Psi_b\right|_{E\approx\pi}
\,\to \, \psib (i\gamma_5\gamma_0)\gamma_5
\Psi_b\,\,(-1)^{t}
\,=\, -\psib i\gamma_0 \Psi_b\,(-1)^{t}.
\ee
The operator $\psib i\gamma_0\Psi_b$, now with low-energy fields,
couples to $0^+$ mesons. (It is $J=0$ because there
is no three-vector
index. It is~$P=+$ because $\Parity\gamma_0\Parity = \gamma_0$ where
$\Parity\equiv\gamma_0$
is the parity operator.)
The full correlator has two components:
\be
\begin{aligned}
G_{55}(t) \,\to\,&
|\langle0|\psib\gamma_5\Psi_b|0^-\rangle|^2\,\e^{-E_-t} \\
& -\, (-1)^t \,
|\langle0|\psib i\gamma_0\Psi_b|0^+\rangle|^2\,\e^{-E_+t}.
\end{aligned}
\ee
The second component, rather unconventionally, oscillates in sign from 
time
step to time step.


\begin{thebibliography}{23}
\bibitem{ratio-paper} %
C.~T.~H.~Davies {\it et al.}  [HPQCD, Fermilab, MILC, UKQCD 
Collaborations],
%
Phys.\ Rev.\ Lett.\  {\bf 92}, 022001 (2004)
[arXiv:hep-lat/0304004].
%

\bibitem{fk-fpi}
C.~Aubin {\it et al.}  [MILC Collaboration],
%
%
Phys.\ Rev.\ D {\bf 70}, 114501 (2004)
[arXiv:hep-lat/0407028].
%

\bibitem{alphas-paper} %
Q.~Mason {\it et al.}  [HPQCD Collaboration],
%
Phys.\ Rev.\ Lett.\  {\bf 95}, 052002 (2005)
[arXiv:hep-lat/0503005].
 %

\bibitem{fd}
C.~Aubin {\it et al.},
 %
 Phys.\ Rev.\ Lett.\  {\bf 95}, 122002 (2005)
 [arXiv:hep-lat/0506030].
 %

\bibitem{mb-paper} %
Q.~Mason, H.~D.~Trottier, R.~Horgan, C.~T.~H.~Davies and G.~P.~Lepage
[HPQCD Collaboration],
%
%
Phys.\ Rev.\ D {\bf 73}, 114501 (2006)
[arXiv:hep-ph/0511160].
%

\bibitem{naik-paper} %
S.~Naik,
%
%
Nucl.\ Phys.\ B {\bf 316}, 238 (1989).
%

\bibitem{doug-papers} %
D.~Toussaint and K.~Orginos  [MILC Collaboration],
%
Nucl.\ Phys.\ Proc.\ Suppl.\  {\bf 73}, 909 (1999)
[arXiv:hep-lat/9809148];
%
%
Phys.\ Rev.\ D {\bf 59}, 014501 (1999)
[arXiv:hep-lat/9805009].
%


\bibitem{lepage1-paper} %
G.~P.~Lepage,
%
Nucl.\ Phys.\ Proc.\ Suppl.\  {\bf 60A}, 267 (1998)
[arXiv:hep-lat/9707026].
%

\bibitem{sinclair-paper}
J.~F.~Lagae and D.~K.~Sinclair,
%
%
Nucl.\ Phys.\ Proc.\ Suppl.\  {\bf 63}, 892 (1998)
[arXiv:hep-lat/9709035];
%
%
%
Phys.\ Rev.\ D {\bf 59}, 014511 (1999)
[arXiv:hep-lat/9806014].
%

\bibitem{lepage2-paper} %
G.~P.~Lepage,
%
%
Phys.\ Rev.\ D {\bf 59}, 074502 (1999)
[arXiv:hep-lat/9809157].
%

\bibitem{hyp-paper} %
A.~Hasenfratz and F.~Knechtli,
%
Phys.\ Rev.\ D {\bf 64}, 034504 (2001)
[arXiv:hep-lat/0103029].
%

\bibitem{nrqcd} %
See, for example, A.~Gray, I.~Allison, C.~T.~H.~Davies, E.~Gulez, 
G.~P.~Lepage, J.~Shigemitsu and M.~Wingate,
%
Phys.\ Rev.\ D {\bf 72}, 094507 (2005)
[arXiv:hep-lat/0507013].
%

\bibitem{ukqcd-w} C.R. Allton et al [UKQCD Collaboration],
PRD65:054502 (2002).

\bibitem{creutz-paper-etc} See, for example,
C.~Bernard, M.~Golterman and Y.~Shamir,
%
arXiv:hep-lat/0610003;
%
M.~Creutz,
%
arXiv:hep-lat/0608020;
%
Y.~Shamir,
%
%
arXiv:hep-lat/0607007;
%
C.~Bernard, M.~Golterman, Y.~Shamir and S.~R.~Sharpe,
%
arXiv:hep-lat/0603027;
%
M.~Creutz,
%
arXiv:hep-lat/0603020.
%


\bibitem{milc-configs}
See, for example, C.~Aubin {\it et al.},
 %
 %
 Phys.\ Rev.\ D {\bf 70}, 094505 (2004)
 [arXiv:hep-lat/0402030];
 %
 C.~Aubin {\it et al.}  [MILC Collaboration],
 %
 Nucl.\ Phys.\ Proc.\ Suppl.\  {\bf 129}, 227 (2004)
 [arXiv:hep-lat/0309088].
 %

\bibitem{tadpole} Tadpole improvement requires dividing every $U_\mu$ 
in the final action by a mean link factor $u_0$ (after expanding 
products of operators so as to remove factors of $U_\nu(x)^\dagger 
U_\nu(x) = 1$); see
G.~P.~Lepage and P.~B.~Mackenzie,
%
Phys.\ Rev.\ D {\bf 48}, 2250 (1993)
[arXiv:hep-lat/9209022].
%


\bibitem{Dslash-spectrum}
E.~Follana, A.~Hart, C.~T.~H.~Davies and Q.~Mason  [HPQCD 
Collaboration],
%
Phys.\ Rev.\ D {\bf 72}, 054501 (2005)
[arXiv:hep-lat/0507011];
%
E.~Follana, A.~Hart and C.~T.~H.~Davies  [HPQCD Collaboration],
%
%
Phys.\ Rev.\ Lett.\  {\bf 93}, 241601 (2004)
[arXiv:hep-lat/0406010].
%

\bibitem{aubin} %
C.~Aubin and C.~Bernard,
 %
 Phys.\ Rev.\ D {\bf 73}, 014515 (2006)
 [arXiv:hep-lat/0510088].
 %

\bibitem{tsukuba}
G.~P.~Lepage,
%
Nucl.\ Phys.\ Proc.\ Suppl.\  {\bf 26}, 45 (1992).
%

\bibitem {schladming}
G.~P.~Lepage, ``Redesigning Lattice QCD,'' published in the proceedings 
of the 35th International School on Nuclear and Particle Physics, 
Schladming, Austria (1996)
%
[arXiv:hep-lat/9607076].

\bibitem{quentin-thesis}
  Q.~J.~Mason,
  %
Cornell University Ph.D. Thesis (2004), UMI-31-14569.

\bibitem{japan-psi}
S.~Aoki {\it et al.},
%
Nucl.\ Phys.\ Proc.\ Suppl.\  {\bf 42}, 303 (1995)
[arXiv:hep-lat/9411058].
%

\bibitem{etac-data}
W.-M.\ Yao et al. [Particle Data Group], J.\ Phys.\ G~\textbf{33}, 1 
(2006).

\bibitem{bodwin-braaten-gpl}
G.~T.~Bodwin, E.~Braaten and G.~P.~Lepage,
%
%
Phys.\ Rev.\ D {\bf 51}, 1125 (1995)
[Erratum-ibid.\ D {\bf 55}, 5853 (1997)]
[arXiv:hep-ph/9407339].
%

\bibitem{positronium-result} The perturbative result for an $\eta_c$ is 
the same, up to an overall factor, as for parapositronium.
See, for example, G. Adkins in \emph{Relativistic, Quantum 
Electrodynamic and Weak Interaction Effects in Atoms}, edited by 
W.\ Johnson et al, AIP Conference Proceedings~\textbf{189}, American 
Institute of Physics (1989).

\bibitem{Follana:2003fe}
E.~Follana, Q.~Mason, C.~Davies, K.~Hornbostel, P.~Lepage and 
H.~Trottier
[HPQCD Collaboration],
%
Nucl.\ Phys.\ Proc.\ Suppl.\  {\bf 129}, 447 (2004)
[arXiv:hep-lat/0311004].
%

\bibitem{Kamleh:2004xk}
W.~Kamleh, D.~B.~Leinweber and A.~G.~Williams,
%
Phys.\ Rev.\ {\bf D70}, 014502 (2004)
[arXiv:hep-lat/0403019].
%

\bibitem{Morningstar:2003gk}
C.~Morningstar and M.~J.~Peardon,
%
Phys.\ Rev.\ D {\bf 69}, 054501 (2004)
[arXiv:hep-lat/0311018].
%

%
%
%
%
%
%
%
%
%
%
%
%
%
%
%
%

\bibitem{lee-sharpe}
W.~Lee and S.~R.~Sharpe,
%
Phys.\ Rev.\ D {\bf 66}, 114501 (2002)
[arXiv:hep-lat/0208018].
%

\bibitem{wingate}
Wingate et al., Phys.\ Rev.\ {\bf D67}, 054505 (2003)
     [hep-lat/0211014].
\end{thebibliography}
\end{document}